\documentclass[lettersize,journal]{IEEEtran}
\usepackage{booktabs,threeparttable}  
\usepackage{adjustbox}       
\usepackage{eqparbox}
\usepackage{multirow}
\usepackage{bm}
\usepackage{url}
\usepackage{tabularx}
\usepackage{booktabs,threeparttable,colortbl}
\usepackage[percent]{overpic} 
\usepackage{graphicx}
\usepackage[switch]{lineno}
\usepackage[ruled,vlined]{algorithm2e}
\usepackage{array}
\usepackage{stfloats}
\usepackage{threeparttable}
\usepackage{verbatim}
\usepackage{url}
\usepackage{cite}
\usepackage{amsmath,amssymb,amsfonts}
 \usepackage[ruled,vlined]{algorithm2e}
 \usepackage[utf8]{inputenc}
\usepackage[ruled,vlined]{algorithm2e}
\usepackage{changepage}
\usepackage{graphicx}
\usepackage{subfigure}
\usepackage{textcomp}
\usepackage{xcolor}
\usepackage{colortbl}
\usepackage{float}
\usepackage{booktabs}
\definecolor{awesome}{rgb}{0.98, 0.66, 0.68}
\usepackage{multirow}
\usepackage{etoolbox}
\usepackage{setspace}
\usepackage{etoolbox}
 
\newcommand{\shift}[2]{\makebox[\linewidth][c]{\hspace{#1}#2}}

\usepackage{graphicx}
\graphicspath{{fig/}}   
\usepackage{cite}
\usepackage{booktabs,threeparttable}  
\usepackage{booktabs}           
\usepackage{amsmath}
\usepackage{threeparttable}     
\usepackage{amsmath,amssymb}
\usepackage{adjustbox}       
\usepackage{graphicx}
\usepackage{eqparbox}
\usepackage{amsmath, amssymb}
\usepackage{tikz}
\usepackage{multirow}
\usetikzlibrary{arrows.meta, positioning}
 \usepackage{graphicx}
\usepackage[caption=false,font=footnotesize]{subfig}

\usepackage{changepage}
\usepackage{bm}
\usepackage{url}
\usepackage{booktabs}
\usepackage{tabularx}
\usepackage{booktabs}
\usepackage{pifont}
\usepackage{booktabs}
\usepackage{siunitx}        
\newcolumntype{Y}{>{\raggedright\arraybackslash}X}    
\usepackage[table]{xcolor} 
\newcommand{\add}[1]{\scriptsize\textcolor{blue}{#1}}

\definecolor{NeonGreen}{HTML}{39FF14} 
\newcommand{\addGre}[1]{\scriptsize\textcolor{NeonGreen}{#1}}

\definecolor{LightRed}{RGB}{255,230,230} 
\definecolor{shallowBlue}{RGB}{173,216,230} 
\definecolor{AlertRed}{RGB}{255,0,0}
\newcommand{\addred}[1]{\scriptsize\textcolor{AlertRed}{#1}}

\def\BibTeX{{\rm B\kern-.05em{\sc i\kern-.025em b}\kern-.08em
    T\kern-.1667em\lower.7ex\hbox{E}\kern-.125emX}}
\usepackage{balance}
\usepackage[table]{xcolor} 

\definecolor{LightRed}{RGB}{255,230,230} 
\definecolor{shallowBlue}{RGB}{173,216,230} 
\newcommand{\bgred}[1]{\colorbox{LightRed}{\strut \textbf{#1}}}
\newcommand{\bgblue}[1]{\colorbox{shallowBlue}{\strut #1}}
\def\BibTeX{{\rm B\kern-.05em{\sc i\kern-.025em b}\kern-.08em
    T\kern-.1667em\lower.7ex\hbox{E}\kern-.125emX}}
\usepackage{balance}

\begin{document}
 
\title {Online Adaptation via Dual-Stage Alignment and Self-Supervision for Fast-Calibration Brain-Computer Interfaces

\thanks{This work was supported in part by the National Key Research and Development Program of China under Grant 2023YFC2415100, in part by the National Natural Science Foundation of China under Grant 62222316, Grant 62373351, Grant 82327801, Grant 62073325, Grant 62303463, in part by the Chinese Academy of Sciences Project for Young Scientists in Basic Research under Grant No. YSBR-104, in part by the Beijing Natural Science Foundation under Grant F252068, Grant 4254107, in part by China Postdoctoral Science Foundation under Grant 2024M763535.} 
  \thanks{Sheng-Bin Duan and Jian-Long Hao are with the School of Information, Shanxi University of Finance and Economics, Taiyuan 030006, China. Email: shengbinduan666@gmail.com.
  }
  \thanks{Xiao-Hu Zhou, Mei-Jiang Gui, Tian-Yu Xiang, Xiao-Liang Xie, and Shi-Qi Liu are with the State Key Laboratory of Multimodal Artificial Intelligence Systems, Institute of Automation, Chinese Academy of Sciences, Beijing 100190, China, and with the School of Artificial Intelligence, University of Chinese Academy of Sciences, Beijing 100049, China.}
  \thanks{Zeng-Guang Hou is with State Key Laboratory of Multimodal Artificial
Intelligence Systems, Institute of Automation, Chinese Academy of Sciences,
Beijing 100190, China, also with the CAS Center for Excellence in Brain
Science and Intelligence Technology, Beijing 100190, China, also with the
School of Artificial Intelligence, University of Chinese Academy of Sciences,
Beijing 100049, China, and also with the Joint Laboratory of Intelligence
Science and Technology, Institute of Systems Engineering, Macau University
of Science and Technology, Taipa, Macao, China.}
\thanks{$\dagger$ Equal contribution: Sheng-Bin Duan, Jian-Long Hao, Tian-Yu Xiang  

$*$ Corresponding author: Xiao-Hu Zhou}

}

\author{
\IEEEauthorblockN{Sheng-Bin Duan$^\dagger$, Jian-Long Hao$^\dagger$, Tian-Yu Xiang$^\dagger$, Xiao-Hu Zhou$^{*}$, \emph{Member, IEEE}, Mei-Jiang Gui, Xiao-Liang Xie, Shi-Qi Liu, Zeng-Guang Hou, \emph{Fellow, IEEE} }
\IEEEauthorblockN{}
}
\markboth{Journal of \LaTeX \ Class Files,~Vol.~xx, No.~x, xx~xxxx} 
{How to Use the IEEEtran \LaTeX \ Templates}

\maketitle

\begin{abstract}
Individual differences in brain activity hinder the online application of electroencephalogram (EEG)-based brain computer interface (BCI) systems. To overcome this limitation, this study proposes an online adaptation algorithm for unseen subjects via dual-stage alignment and self-supervision. The alignment process begins by applying Euclidean alignment in the EEG data space and then updates batch normalization statistics in the representation space. Moreover, a self-supervised loss is designed to update the decoder. The loss is computed by soft pseudo-labels derived from the decoder as a proxy for the unknown ground truth, and is calibrated by Shannon entropy to facilitate self-supervised training. Experiments across five public datasets and seven decoders show the proposed algorithm can be integrated seamlessly regardless of BCI paradigm and decoder architecture. In each iteration, the decoder is updated with a single online trial, which yields average accuracy gains of 4.9\% on steady-state visual evoked potentials (SSVEP) and 3.6\% on motor imagery. 
These results support fast-calibration operation and show that the proposed algorithm has great potential for BCI applications. The code will be available upon acceptance.

\end{abstract} 

\begin{IEEEkeywords}
BCI, EEG, Test-Time Adaptation
\end{IEEEkeywords}

\section{Introduction}

EEG-based BCI systems are attractive for research and practical deployment owing to their non-invasive and portable nature~\cite{bouton2016restoring,metzger2023high}. Among these, steady-state visually evoked potentials (SSVEP)~\cite{cheng2002design} and motor imagery~\cite{kawala2021summary} are widely used because they yield stable, physiologically grounded signatures in EEG recordings. In SSVEP, fixation on a flickering visual stimulus elicits responses predominantly over the occipital cortex at the stimulus frequency and its harmonics, providing robust spectral features for decoding~\cite{cheng2002design}. In motor imagery, imagined movement induces event-related desynchronization/synchronization (ERD/S) in sensorimotor rhythms~\cite{pfurtscheller1979evaluation,xiang2024learning,wu2014probabilistic,wang2023sparse,xiang2024learning1}, characterized by a power decrease during imagery (ERD) and a rebound upon return to rest (ERS).

Although SSVEP and motor imagery paradigms generally yield stable features in EEG signals, substantial inter-subject variability persists from macroscale cortical anatomy to microscopic functional organization~\cite{fischl2008cortical,van2004surface,cox2003functional, guntupalli2016model, haxby2011common}. This variability complicates group-level analyses, hinders the transfer of models tuned to individual-specific features, and degrades generalization performance on unseen subjects~\cite{li2023t,maiseli2023brain,wang2025inter}.

To mitigate performance degradation arising from individual differences, existing methods generally align target and source domains at the data-~\cite{zanini2017transfer,he2019transfer,sun2016return} or representation-level~\cite{long2015learning,long2017deep,liang2021domain}. Data-level alignment acts directly on the input data by making the source and target input distributions similar before feature extraction, via data-space transformations, resampling, distribution mapping or translation~\cite{zanini2017transfer,he2019transfer,sun2016return}. Representation-level alignment operates on deep features across domains via adversarial training, domain-invariant mappings, or statistical divergence minimization between source and target features~\cite{long2015learning,long2017deep,liang2021domain}. However, both categories typically require the prior collection of sufficient labeled target-domain data to learn alignment mappings. Such data acquisition is time-consuming and labor-intensive, imposes burdens on users, and reduces the practicality of online BCI.

Test-time adaptation (TTA)~\cite{liang2021source} is one solution to the above challenges~\cite{li2018adaptive,wimpff2025fine,iwasawa2021test,wang2020tent,ma2024improved,zhao2023delta,liu2021ttt++,gandelsman2022test}. TTA leverages unlabeled target-domain data during inference to update model parameters or normalization statistics and mitigating degradation under distribution shifts. Although TTA has achieved notable success in other fields~\cite{sun2020test}, applications to EEG-based BCI systems remain limited. Notable examples include T-TIME by Li et al.~\cite{li2023t}, which minimizes conditional entropy with adaptive marginal distribution regularization to enable efficient and stable online adaptation and decoding, and Wimpff et al.~\cite{wimpff2025fine}, which combines Euclidean alignment with online updates of batch normalization statistics to improve robustness and generalization in non-stationary environments.

However, T-TIME~\cite{li2023t} requires accumulating multiple EEG trials before model update in the initial stage. As each trial typically lasts several seconds, this induces calibration latency that degrades online responsiveness and increases subjects' burden. Wimpff et al.~\cite{wimpff2025fine} perform alignment at the data and representation levels without updating model parameters, which limits generalization to unseen subjects. In summary, TTA for online BCI systems still faces two central challenges:
\begin{enumerate}
  \item \textbf{Inter-subject variability:} Individual differences induce source-target mismatches that degrade performance on unseen subjects.
  \item \textbf{Fast-calibration, efficient adaptation:} Reducing multiple trial accumulation while enabling stable online updates to preserve online responsiveness and reduce subject burden.
\end{enumerate}

To address these challenges, a dual-stage alignment is employed for both the input data-level and the representation-level. Subsequently, the model parameters are updated to adapt to the target-domain distribution using a self-supervised loss. Both dual-stage alignment and the self-supervised optimization encourage the algorithm to support single-trial calibration, reducing calibration latency. The main contributions of this study are as follows:

\begin{itemize}
\item A plug-and-play online adaptation algorithm with dual-stage alignment and self-supervision is proposed for EEG-based BCI systems, enabling fast calibration via single-trial update.
\item A dual-stage alignment strategy applies Euclidean alignment to the input data and then adjusts batch normalization statistics for the representations. 
\item A self-supervised loss is computed using soft pseudo-labels as a proxy for unavailable ground truth and calibrated by Shannon entropy to facilitate stable self-supervised training.
\item Extensive experiments on five public datasets and seven backbone networks demonstrate that the proposed algorithm delivers significant performance improvements in online settings for two widely used BCI paradigms, achieving average accuracy gains of 4.9\% for SSVEP and 3.6\% for motor imagery tasks.
\end{itemize}

\section{Related Works}
 
\subsection{Offline Alignment Methods}

Offline alignment methods address source–target domain distribution shifts during data pre-processing or model training by introducing adjustments in either the data space or the representation space. Once the model is deployed for online testing, its parameters remain fixed. Such methods are commonly divided into input data-level alignment and representation-level alignment.

\subsubsection{Data-Level Distribution Alignment}

Data-level distribution alignment methods mitigate domain shifts directly by processing input data via transformations or resampling/weighting strategies. Input transformation approaches adjust source–target statistics to align distributions. Riemannian alignment~\cite{zanini2017transfer,fathy2016discriminative} and Euclidean alignment~\cite{he2019transfer,bakas2025latent,junqueira2024systematic} align the covariance structure in the corresponding spaces. Correlation alignment~\cite{sun2016return} whitens source features and then re-colors them so that second-order statistics match the target covariance. Resampling and importance weighting approaches reweight or select samples to better reflect the target distribution. Source samples more similar to the target domain receive higher weights, whereas dissimilar ones are down-weighted; equivalently, the sampling distribution can be adjusted to balance source and target~\cite{shimodaira2000improving,huang2006correcting,sugiyama2007covariate}. In practice, rare target-domain patterns may be oversampled~\cite{zhao2022t,chawla2002smote}, while mismatched source samples are undersampled. By aligning the training distribution to the target domain, these methods alleviate performance degradation due to marginal distribution shifts. Examples include importance weighting and its practical variants~\cite{huang2006correcting,sugiyama2007covariate,zhao2022t,chawla2002smote}.

Data-level alignment methods have reduced the impact of distribution shifts and improved generalization. However, they typically adopt a batch/offline assumption: target-domain data are fully available before inference and can be revisited repeatedly. This assumption poses challenges for online BCI systems, where samples arrive sequentially.


\subsubsection{Representation-Level Distribution Alignment}
Representation-level distribution alignment approaches operate on latent features, aiming to learn domain-invariant representations by explicitly reducing the discrepancy between source and target data in the representation spaces. For instance, the deep adaptation network~\cite{long2015learning} aligns deep feature activations using a kernel-based maximum mean discrepancy. The joint adaptation network~\cite{long2017deep} extends this strategy by aligning both marginal and conditional distributions to achieve class-level feature matching. Subsequently, Liang et al.~\cite{liang2021domain} introduce an auxiliary classifier to generate the pseudo labels for target samples, improving the reliability and stability of alignment. More recently, Zhang et al.~\cite{zhang2022self} employ self-supervised contrastive learning to align time-domain and frequency-domain embeddings, enabling the model to experience a broader range of distributional variations during pre-training and thus enhancing robustness to unseen domains. 

Although these methods effectively mitigate the effects of distribution shifts at the representation level and demonstrate strong performance across a variety of tasks, the alignment process remains dependent on labeled data. In practice, additional labeled target-domain data are often required for supervised tuning to further alleviate distribution mismatches. However, labeled target-domain data are generally unavailable in online BCI, which can complicate supervised tuning.

\subsection{Online Adaptation Techniques}

Online adaptation techniques update model parameters during testing by exploiting unlabeled target-domain data or model predictions, allowing the model to continuously preserve robustness under dynamic conditions~\cite{voigtlaender2017online}.

\subsubsection{Unsupervised Domain Adaptation}

The objective of unsupervised domain adaptation is to transfer knowledge from a labeled source domain to an unlabeled target domain, mitigating distributional discrepancies and yielding a model that generalizes to the target domain~\cite{ganin2015unsupervised, liu2022deep, chen2025unsupervised}. A stricter variant, source-free UDA (SFUDA)~\cite{fang2024source}, assumes access only to a pre-trained source model and unlabeled target data; adaptation proceeds by updating the source model using the target data, potentially in an online manner, to compensate for distribution shifts during deployment. Liang et al. propose source hypothesis transfer, which enhances target domain representations via information maximization through a self-supervised training paradigm, promoting feature consistency and improving  performance~\cite{liang2021source, li2021imbalanced}. Collectively, these approaches report reduced domain discrepancies and improved target domain generalization.

However, in online BCI systems, prevailing methods typically require buffer unlabeled target data to enable stable, iterative optimization. Such batch-accumulation schemes introduce waiting periods and calibration latency in BCI systems.

\subsubsection{Test‑Time Adaptation and Test-Time Training}

Another category of online adaptation comprises test-time adaptation and test-time training~\cite{sun2020test}. These approaches exploit unlabeled test data during inference to improve performance on the target domain. Existing approaches can be grouped into three categories: \textbf{(i)} adjusting model predictions during testing using target-domain statistics without backpropagation (e.g., prototype construction or updating batch normalization statistics)~\cite{li2018adaptive,wimpff2025fine,iwasawa2021test,lee2025prototypical}; \textbf{(ii)} fine-tuning parameters during testing through a tailored unsupervised loss on target data to enhance prediction robustness (e.g., entropy minimization, pseudo-labeling, and output-diversity regularization)~\cite{wang2020tent,ma2024improved,zhao2023delta,li2023t}; and \textbf{(iii)} optimizing a self-supervised loss on target-domain inputs (e.g., reconstruction or contrastive learning), where the associated loss drives adaptation on unlabeled data and helps maintain stability under distribution shifts~\cite{sun2020test,liu2021ttt++,gandelsman2022test}.

However, these methods often fail to satisfy the low calibration latency requirements of online BCI systems. When each parameter update requires aggregating data across multiple trials, the ensuing delay degrades update efficiency and increases the burden on the participants.

\begin{figure*}
    \centering
    \includegraphics[width=1\linewidth]{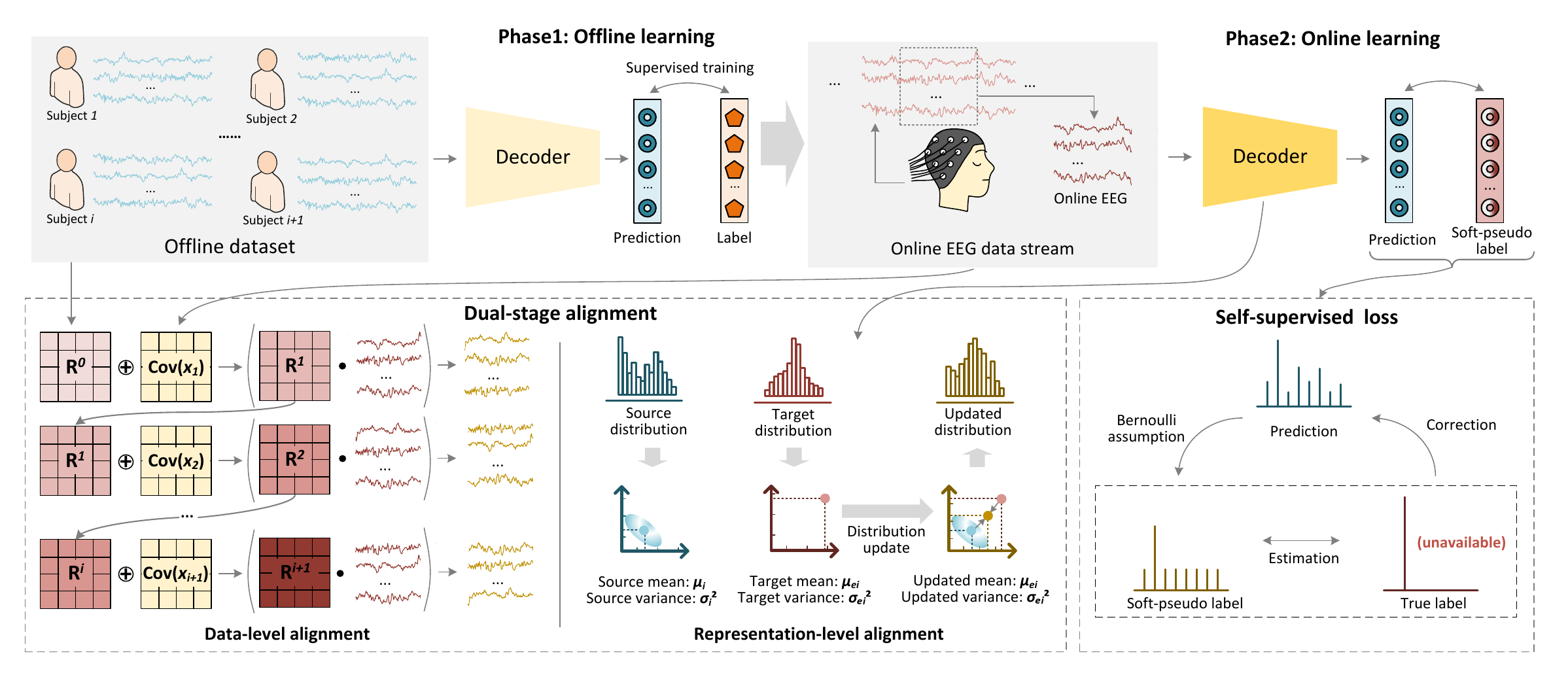}
    \caption{In the offline phase, the source-domain EEG signals are first aligned using Euclidean alignment, followed by supervised training to obtain a pre-trained decoder. In the testing phase, each new trial undergoes a dual-stage alignment: Stage~1 applies online Euclidean alignment to align the input EEG signals into a unified Euclidean space, and Stage~2 aligns the intermediate representations by updating the batch normalization statistics. The decoder is updated through a self-supervised loss computed by soft pseudo-labels as a proxy for unavailable ground truth.}
    \label{fig:frame}
\end{figure*}

\section{Methods}

A plug-and-play online adaptation algorithm that supports single-trial calibration is realized via dual-stage alignment in the EEG data and representation spaces, combined with a self-supervised loss for updating the decoder (shown in Fig.~\ref{fig:frame}).

\subsection{Problem Definition}
 
Consider a \(K\)-class EEG decoding problem. The training dataset \(\mathcal{D}\) is:
\begin{equation}
\label{eq:dataset}
\mathcal{D}=\{(x_i^{\mathrm{train}},y_i^{\mathrm{train}})\}_{i=1}^{N},\quad x_i^{\mathrm{train}}\in\mathbb{R}^{C\times T},
\end{equation}
where \(x_i^{\mathrm{train}}\) denotes the \(i\)th EEG trial with \(C\) channels and \(T\) time points, \(y_i^{\mathrm{train}}\) is the corresponding class label, and \(N\) is the number of training trials.

The standard cross-entropy ($\mathcal{L}_{\mathrm{CE}}$) loss is used for the initial offline training:
 \begin{equation}
\mathcal{L}_{\mathrm{CE}}
= -\frac{1}{B}\sum_{i=1}^B \log q\bigl(\hat{Y}^\mathrm{train}_i = y^\mathrm{train}_i \mid x^\mathrm{train}_i\bigr)
\end{equation}
where $\hat{Y}_i^{\mathrm{train}}\in\{1,\dots,K\}$ denotes the random variable for the predicted class of the model on the $i$th trial $x_i^{\mathrm{train}}$. \(q(\hat{Y}_i^{\mathrm{train}}=y^\mathrm{train}_i\mid x^\mathrm{train}_i)\) is the probability of the true class predicted by the model, and \(B\) denotes the batch-size.

After offline training yields parameters \(\theta_{0}\), the model is adapted on a stream of test trials \(x_i^{\mathrm{test}}\). For each trial, the decoder performs EEG data-level and representation-level alignment to the target domain and updates the parameters by minimizing a self-supervised loss \(\mathcal{L}_{\mathrm{test}}(x_i^{\mathrm{test}};\theta_{i-1})\) with an online learning rate \(\eta\). And the adaptation process employs a single-trial update scheme: parameters are updated immediately as each test trial arrives, enabling online adaptation.
\begin{equation}
\label{eq:theta_update}
\theta_i \leftarrow \theta_{i-1} - \eta\,\nabla_{\theta_{i-1}}\mathcal{L_\mathrm{test}}(x_i^{\mathrm{test}};\,\theta_{i-1})
\end{equation}
where \(\nabla_{\theta_{i-1}}\) is the gradient computed solely from the current trial \(x_i^{\mathrm{test}}\); no additional labels are required at test time.

\subsection{Dual-Stage Distribution Alignment}

\subsubsection{Input Data-Level Distribution Alignment via Euclidean Alignment}
\paragraph{Offline Phase}
In the offline phase, Euclidean alignment computes the mean covariance of the source domain to define a reference matrix, whose inverse square root whitens the EEG signals in the training set and maps all trials to a unified Euclidean space. 

Let \(\{x_i^{\mathrm{train}}\}_{i=1}^{N}\) denote \(N\) training trials with \(x_i^{\mathrm{train}}\in\mathbb{R}^{C\times T}\). For each trial, the channel-wise covariance is
\begin{equation}
\label{eq:ea_cov}
\mathrm{cov}(x^\mathrm{train}_i) 
\;=\; \frac{1}{T - 1} \sum_{t=1}^{T} 
\Bigl[x^\mathrm{train}_i - \bar{x}^\mathrm{train}_i\Bigr]\,\Bigl[x^\mathrm{train}_i - \bar{x}^\mathrm{train}_i\Bigr]^\top
\end{equation}
where $\mathrm{cov}(\cdot) $ denotes the operator for computing covariance matrix, and \(\bar{x}_i^{\mathrm{train}}\in\mathbb{R}^{C}\) is the channel-wise mean over time. The reference matrix $\bm{R}^{\mathrm{train}}$ is the average covariance:
\begin{equation}
\label{eq:ea_r}
\bm{R}^{\mathrm{train}}=\frac{1}{N}\sum_{i=1}^{N}\mathrm{cov}\!\left(x_i^{\mathrm{train}}\right)
\end{equation}
Its inverse square root follows from the eigen-decomposition:
\begin{equation}
\begin{aligned}
\label{eq:whiling}
&(\bm{R}^{\mathrm{train}})^{-\tfrac12}
 = \bm{U}^\mathrm{train}\,(\Lambda{^\mathrm{train}})^{-\tfrac12}\,(\bm U^\mathrm{train})^\top
\end{aligned}
\end{equation}
where \({\bm U^\mathrm{train}}\) and \(\Lambda^\mathrm{train}\) denote the orthogonal eigenvector matrix and the diagonal eigenvalue matrix of \(\bm{R}^{\mathrm{train}}\), respectively.
Before the offline training phase, all training data are aligned into a unified Euclidean space via:
\begin{equation}
\label{eq:ea_align_train}
\tilde{x}^\mathrm{train}_i
\;=\; \bigl(\bm{R}^{\mathrm{train}}\bigr)^{-\tfrac12}\,x^\mathrm{train}_i,\quad i=1,2,\dots,N
\end{equation}
where \(\tilde{x}^\mathrm{train}_i\) denotes the aligned training data.
\paragraph{Online Phase}

In the online phase, upon receiving a test trial \(x_i^{\mathrm{test}}\), its covariance is computed and the reference matrix is updated by a weighted average of the previous reference and the current trial covariance; the updated reference then whitens the current trial:
\begin{equation}
\label{eq:reference_update}
\bm{R}^{i} =
\begin{cases}
\bm{R}^{\mathrm{train}}, & i = 0,\\[6pt]
\displaystyle\frac{n_{i-1}\,\bm{R}^{i-1} + \omega\,\mathrm{cov}(x^\mathrm{test}_i)}{n_{i-1} + \omega}, & i \ge 1
\end{cases}
\end{equation}
where \(n_{i-1}\) is the number of trials incorporated prior to trial \(i\) (training plus already processed tests) and \(\omega\) is a hyper-parameter controlling the influence of the current trial. The aligned test trial is:
\begin{equation}
\label{eq:eatest_align}
\tilde{x}^\mathrm{test}_i 
\;=\; \bigl(\bm R^{i}\bigr)^{-\tfrac12}\,x^\mathrm{test}_i
\end{equation}

This continuous single-trial update both maintains alignment to the source covariance structure and reduces calibration latency from batch accumulation.

\subsubsection{Intermediate Representations-Level Distribution Alignment by Batch Normalization Statistics Updating}

Despite aligning data space distributions through Euclidean alignment, discrepancies can persist in intermediate representations~\cite{li2016revisiting}. To mitigate this, batch normalization statistics are updated online so that the feature distribution adapts continuously to each target-domain trial. For a batch normalization layer, the outputs are
\begin{equation}
\begin{aligned}
\label{eq:BN_update}
&\tilde{e}_i = \frac{e_i - \mu_{i}}{\sqrt{\sigma_{i}^{2} + \epsilon}},\\\
&\quad z_i = \gamma \, \tilde{e}_i + \beta
\end{aligned}
\end{equation}
where \(\mu_i\) and \(\sigma_i^{2}\) are the running mean and variance at trial \(i\); \(\gamma\) and \(\beta\) are the learnable affine parameters; \(e_i\) and \(\tilde{e}_i\) denote the input to batch normalization and its normalized output, respectively; \(z_i\) is the batch normalization output; and \(\epsilon\) ensures numerical stability.

Single-trial updates of the batch normalization statistics use an exponential moving average~\cite{van2017neural}:
\begin{equation}
\label{eq:BN_align}
\left\{
\begin{aligned}
\mu_{i} &= (1-\alpha)\,\mu_{i-1} + \alpha\,\mu_{e_i}, \\
\sigma_{i}^{2} &= (1-\alpha)\,\sigma_{i-1}^{2} + \alpha\,\sigma_{e_i}^{2} + \alpha(1-\alpha)\,\bigl(\mu_{e_i} - \mu_{i-1}\bigr)^{2}
\end{aligned}
\right.
\end{equation}
where \(\mu_{e_i}\) and \(\sigma_{e_i}^{2}\) are the mean and variance of the current trial’s representations, and \(\alpha\) controls the contribution of the current trial. This per trial exponential moving average~(EMA) strategy aligns intermediate statistics to the evolving target distribution without backpropagation.












\subsection{Soft Pseudo-Label Self-Supervised Loss with Shannon Entropy Regularization}

At test time, ground truth labels are unavailable, so the cross-entropy loss \(\mathcal{L}_{\mathrm{CE}}\) cannot be evaluated directly. A common way is using hard pseudo labels: each trial is assigned to the class with the highest predicted probability, and this one-hot assignment is used as a proxy label for optimization~\cite{lee2013pseudo}. However, hard pseudo-labels can reinforce erroneous predictions and introduce label noise, producing biased supervision and degraded performance~\cite{arazo2020pseudo}.

To reduce the noise induced by hard pseudo-labels, the model’s predictive probability vector is used as a soft pseudo-label, avoiding discretization into a one-hot target. Because Shannon entropy is defined on probability distributions, an entropy objective ($\mathcal{L}_{\mathrm{Ent}}$) can be applied directly to these predictions during test-time adaptation~\cite{wang2020tent}. Specifically,
\begin{equation}
\label{eq:ent_def}
\mathcal{L}_{\mathrm{Ent}}
= -\sum_{k=1}^{K} q\bigl(\hat Y_i^{\mathrm{test}} = k \mid x_i^{\mathrm{test}}\bigr)
\log q\bigl(\hat Y_i^{\mathrm{test}} = k \mid x_i^{\mathrm{test}}\bigr)
\end{equation}
where $q(\hat Y_i^{\mathrm{test}} = k \mid x_i^{\mathrm{test}})$ denotes the predicted probability of class $k$ for test trial $x_i^{\mathrm{test}}$. Minimizing \(\mathcal{L}_{\mathrm{Ent}}\) encourages confident, self-consistent predictions without requiring ground truth labels, serving as a label-free proxy to the cross-entropy objective when true labels are unavailable~\cite{wang2020tent}.
 
Although \(\mathcal{L}_{\mathrm{Ent}}\) can serve as a proxy for \(\mathcal{L}_{\mathrm{CE}}\) during unlabeled testing, a gap $\Delta$ remains:
\begin{equation}
    \label{eq:delta}
    \Delta \;=\; \mathcal{L}_{\mathrm{CE}} \;-\; \mathcal{L}_{\mathrm{Ent}}
\end{equation}

Because \(\Delta\) depends on the unknown supervised term \(\mathcal{L}_{\mathrm{CE}}\), an accuracy-calibrated label model is adopted to approximate it by the following assumption.
 
\noindent\textbf{Assumption:} 
For each test trial, the target label follows a categorical distribution that places probability \(A_{\mathrm{test}}\) on the correct class and distributes the remaining mass uniformly across the other classes (Bernoulli distribution). Accordingly, define the proxy label distribution as:
\begin{equation}
  \hat p^{\mathrm{test}}\bigl(\hat Y_i^{\mathrm{test}} = k \mid x_i^{\mathrm{test}}\bigr)
  =
  \begin{cases}
    A_{\mathrm{test}}, & k = y_i^{\mathrm{test}}, \\[4pt]
    \dfrac{1 - A_{\mathrm{test}}}{K - 1}, & k \neq y_i^{\mathrm{test}}
  \end{cases}
  \label{eq:appendix_true}
\end{equation}
where \(y_i^{\mathrm{test}}\) is the (unobserved) true class of \(x_i^{\mathrm{test}}\). Since \(A_{\mathrm{test}}\) is unknown a priori, it is approximated by the validation accuracy \(A_{\mathrm{val}}\); Appendix~A provides a justification for this substitution.

The predicted class for \(x_i^{\mathrm{test}}\) is:
\begin{equation}
  y_i^* = \arg\max_{k}\;q\bigl(\hat Y_i^{\mathrm{test}} = k \mid x_i^{\mathrm{test}}\bigr)
  \label{eq:pred_class}
\end{equation}
using \(A_{\mathrm{val}}\) and the predicted class ($y_i^*$), define the accuracy-calibrated soft pseudo-label distribution:
\begin{equation}
  \hat p^{\mathrm{val}}\bigl(\hat Y_i^{\mathrm{test}} = k \mid x_i^{\mathrm{test}}\bigr)
  =
  \begin{cases}
    A_{\mathrm{val}}, & k = y_i^*, \\[4pt]
    \dfrac{1 - A_{\mathrm{val}}}{K - 1}, & k \neq y_i^*
  \end{cases}
  \label{eq:pseudo_label}
\end{equation}
where $\hat p^{\mathrm{val}}\bigl(\hat Y_i^{\mathrm{test}} = k \mid x_i^{\mathrm{test}}\bigr)$ denotes the soft pseudo‑label distribution computed using \(A_{\mathrm{val}}\). The corresponding soft cross-entropy proxy $\hat{\mathcal{L}}_{\mathrm{CE}}^{\mathrm{val}}$ is defined as:
\begin{equation}
  \hat{\mathcal{L}}_{\mathrm{CE}}^{\mathrm{val}}
  = -\sum_{k=1}^{K}
    \hat p^{\mathrm{val}}\bigl(\hat Y_i^{\mathrm{test}} = k \mid x_i^{\mathrm{test}}\bigr)
    \,\log
    q\bigl(\hat Y_i^{\mathrm{test}} = k \mid x_i^{\mathrm{test}}\bigr)
  \label{eq:ce_pseudo}
\end{equation}

In practice, a mismatch between \(A_{\mathrm{val}}\) and the unknown \(A_{\mathrm{test}}\) induces bias in \(\hat{\mathcal{L}}_{\mathrm{CE}}^{\mathrm{val}}\). To mitigate this effect, a hyper-parameter \(\lambda\) balances entropy minimization and the calibrated cross-entropy, yielding the self-supervised objective:
\begin{equation}
\label{eq:total_loss_final}
\mathcal{L}_{\mathrm{test}} 
= \mathcal{L}_{\mathrm{Ent}}
+ \lambda\bigl(\hat{\mathcal{L}}_{{\mathrm{CE}}}^\mathrm{val} - \mathcal{L}_{\mathrm{Ent}}\bigr)
\end{equation}
where the hyper-parameter \(\lambda\) controls the degree of correction.

\section{Experiments}

The experiments evaluate the following aspects of the proposed method:
\begin{itemize}
    \item Can the algorithm be applied to different datasets?
    \item Can the algorithm be applied across diverse BCI paradigms?
    \item Is the algorithm plug-and-play with a range of EEG classification networks?
    \item Can the algorithm deliver consistent performance improvements across evaluation categories?
    \item Does the proposed algorithm outperform other TTA algorithms? 
\end{itemize}
\subsection{Experimental Setup and Details}
All experiments are implemented by PyTorch 1.13.0 and follow a leave-one-subject-out (LOSO) cross-validation protocol, in which one subject is held out as the test set and the remaining subjects form the training set fold. Within the training set, for every subject, the last 20\% of trials are used for validation and the remaining 80\% for training. 

For the SSVEP paradigm, baseline models strictly followed the configurations reported in the original papers~\cite{ravi2020comparing,zhao2021filter}. For the motor imagery paradigm, detailed hyperparameter settings are provided in Table~\ref{tab:mi-hyperparams}; hyperparameters are selected based on validation performance within the LOSO protocol. For SSVEP, each trial is segmented into non-overlapping 1-second windows for both training and testing following the previous studies~\cite{ravi2020comparing,zhao2021filter}, and for the motor imagery paradigm full trial length is used.

The remaining hyperparameters are fixed as follows: \(\omega=500\); batch normalization EMA weight \(\alpha=0.7\) for both SSVEP and motor imagery; $\epsilon = 3\times10^{-5}$; and $\lambda > 1$.


\begin{table}[!t]
  \centering
  \begin{threeparttable}
    \caption{Hyper-parameter Settings for Baseline Models in Motor Imagery Paradigm.}
    \label{tab:mi-hyperparams}
    \begin{tabular}{llll}
      \toprule
      \textbf{Model} & \textbf{Dataset} & \textbf{BS} & \textbf{LR} \\
      \midrule
      \multirow{3}{*}{EEGNet~\cite{lawhern2018eegnet}} 
        & PhysioNet~\cite{schalk2004bci2000}       & 200 & $1\times10^{-4}$ \\
        & Moritz~\cite{grosse2009beamforming}      & 200 & $1\times10^{-4}$ \\
        & Zhou2016~\cite{zhou2016fully}            & 200 & $1\times10^{-4}$ \\
      \midrule
      \multirow{3}{*}{DeepConvNet~\cite{schirrmeister2017deep}}
        & PhysioNet       & 200 & $1\times10^{-4}$ \\
        & Moritz          & 200 & $3\times10^{-5}$ \\
        & Zhou2016        & 200 & $1\times10^{-4}$ \\
      \midrule
      \multirow{3}{*}{EEGConformer~\cite{song2022eeg}}
        & PhysioNet       & 100 & $5\times10^{-6}$ \\
        & Moritz          & 100 & $5\times10^{-6}$ \\
        & Zhou2016        & 200 & $1\times10^{-4}$ \\
      \bottomrule
    \end{tabular}
    \begin{tablenotes}
      \footnotesize
      \item[1] BS denotes batch-size, LR denotes learning rate, and the motor imagery training epochs are set to 600.
    \end{tablenotes}
  \end{threeparttable}
\end{table}

\begin{table*}[!htb]\small
  \centering
  \renewcommand{\arraystretch}{1.2}
   \caption{Inter‐subject classification accuracies (\%) on the SSVEP Benchmark and 12JFPM datasets.}
  \label{tab:combined}
  \begin{threeparttable}
 
  \begin{tabular}{
    p{2.3cm}<{\raggedright\arraybackslash}  
    p{1.3cm}<{\centering} p{1.2cm}<{\centering} p{1.3cm}<{\centering} p{1.2cm}<{\centering}  
    p{1.3cm}<{\centering} p{1.2cm}<{\centering} p{1.3cm}<{\centering} p{1.2cm}<{\centering}  
    p{1.5cm}<{\centering}  
  }
  \toprule
 
  \multirow{2.5}{*}{\textbf{Method}}
    & \multicolumn{4}{c}{\textbf{Benchmark}}
    & \multicolumn{4}{c}{\textbf{12JFPM}}
    & \multirow{2.5}{*}{\textbf{Mean}} \\
  \cmidrule(lr){2-5} \cmidrule(lr){6-9}
    &\rotatebox{0}{\textbf{M1}} & \rotatebox{0}{\textbf{M2}} & \rotatebox{0}{\textbf{M3}} & \rotatebox{0}{\textbf{M4}}
      & \rotatebox{0}{\textbf{M1}} & \rotatebox{0}{\textbf{M2}} & \rotatebox{0}{\textbf{M3}} & \rotatebox{0}{\textbf{M4}} &  \\
  \midrule
 
  Baseline
      & 59.68 & 70.15 & 63.27 & 73.53
      & 70.11 & 81.02 & 73.46 & 83.37 & 71.82 \\
 \midrule
  AdaBN
      & 58.15\add{↓1.53} & 69.25\add{↓0.90} & 61.69\add{↓1.58} & 72.45\add{↓1.08}
      & 71.33\addred{↑1.22} & \cellcolor{shallowBlue}82.50\addred{↑1.48} & 73.76\addred{↑0.30}& \cellcolor{LightRed}\textbf{83.71}\addred{↑0.34} & 71.61\add{↓0.21} \\
  AdaBN (W/ EA)
      & 63.42\addred{↑3.74} & 69.65\add{↓0.50} & 72.63\addred{↑9.36} & 75.86\addred{↑2.33}
      & 74.28\addred{↑4.17} & 81.58\addred{↑0.56} & 73.55\addred{↑0.09} & 82.73\add{↓0.64} & 74.21\addred{↑2.39} \\
      
  Tent
      & 57.44\add{↓2.24} & 67.16\add{↓2.09} & 61.36\add{↓1.91} & 70.73\add{↓2.80}
      & 70.36\addred{↑0.25} & 82.19\addred{↑1.17} & 70.03\add{↓3.43} & 80.71\add{↓2.66} & 70.00\add{↓1.82} \\  

  Tent (W/ EA)
      & 63.17\addred{↑3.49} & 66.83\add{↓3.32} & 72.69\addred{↑9.42} & 74.18\addred{↑0.65}
      & 71.64\addred{↑1.53} & 79.00\add{↓2.02} & 72.03\add{↓1.43} & 80.85\add{↓2.52} & 72.55\addred{↑0.73} \\
  T3A 
      & 51.41\add{↓8.27} & 49.64\add{↓20.51} & 58.46\add{↓4.81} & 66.86\add{↓6.67}
      & 70.11\addGre{~~~0.00} & 80.96\add{↓0.06} & 74.91\addred{↑1.45} & 82.93\add{↓0.44} & 66.91\add{↓4.91} \\
  T3A (W/ EA)
      & 55.28\add{↓4.40} & 50.28\add{↓19.87} & 67.41\add{↑4.14} & 70.36\add{↓3.17}
      & \cellcolor{shallowBlue}74.32\addred{↑4.21} & 81.07\addred{↑0.05} & \cellcolor{shallowBlue}74.97\addred{↑1.51} & 82.69\add{↓0.68} & 69.55\add{↓2.27} \\

  T-TIME$^{\star}$
      &  65.34\addred{↑5.66} & 72.15\addred{↑2.00} &  73.57\addred{↑10.30} & \cellcolor{LightRed}\textbf{80.60}\addred{↑7.07}
      & 74.18\addred{↑4.07} & 82.15\addred{↑1.13} & 74.95\addred{↑1.49} & \cellcolor{shallowBlue}82.99\add{↓0.38} & \cellcolor{shallowBlue}75.74\addred{↑3.92} \\
  T-TIME$^{\star\star}$
      & \cellcolor{shallowBlue}66.82\addred{↑7.14} & \cellcolor{shallowBlue}73.80\addred{↑3.65} & \cellcolor{shallowBlue}74.77\addred{↑11.50} & 77.23\addred{↑3.70}
      & 73.68\addred{↑3.57} & 79.40\add{↓1.62} & 74.42\addred{↑0.96} & 82.94\add{↓0.43} & 75.38\addred{↑3.56} \\
  \midrule
  Ours
      & \cellcolor{LightRed}\textbf{67.18}\addred{↑7.50} & \cellcolor{LightRed}\textbf{74.46}\addred{↑4.31} & \cellcolor{LightRed}\textbf{74.98}\addred{↑11.71} & \cellcolor{shallowBlue}80.31\addred{↑6.78}
      & \cellcolor{LightRed}\textbf{74.89}\addred{↑4.78} & \cellcolor{LightRed}\textbf{83.47}\addred{↑2.45} & \cellcolor{LightRed}\textbf{75.51}\addred{↑2.05} & \cellcolor{shallowBlue}82.99\add{↓0.38} & \cellcolor{LightRed}\textbf{76.72}\addred{↑4.90} \\
  \bottomrule
  \end{tabular}
  \begin{tablenotes}
      \footnotesize
          \item[1] M1–M4 denote CNN-M, CNN-C, FBCNN-M, and FBCNN-C, respectively. 
      
          \item[2] The results compare the proposed algorithm with four TTA methods: AdaBN, Tent, T3A, and T‑TIME ($^{\star}$single-trial update, $^{\star\star}$batch update with batch-size = 8) on 1‑second time window data. 
          
          \item[3] Listed are the performance of AdaBN, Tent, and T3A with Euclidean alignment (W/ EA).
          
          \item[4] The best result per column is in\bgred{red}; the second best is in\bgblue{blue}.  
  \end{tablenotes}
  \end{threeparttable}
\end{table*}

\begin{table*}[!htb]\small
  \centering
  \renewcommand{\arraystretch}{1.2}
  \caption{Inter‐subject classification accuracies (\%) on the motor imagery PhysioNet, Moritz, and Zhou2016 datasets.}
 \label{tab:mi_combined}
  \begin{threeparttable}
  \begin{tabular}{
    p{2.2cm}<{\raggedright\arraybackslash}
    p{1.1cm}<{\centering} p{1.1cm}<{\centering} p{1.1cm}<{\centering}
    p{1.1cm}<{\centering} p{1.1cm}<{\centering} p{1.1cm}<{\centering}
    p{1.1cm}<{\centering} p{1.1cm}<{\centering} p{1.1cm}<{\centering}
    p{1.4cm}<{\centering}
  }
  \toprule
  \multirow{2.5}{*}{\textbf{Method}}
    & \multicolumn{3}{c}{\textbf{PhysioNet}}
    & \multicolumn{3}{c}{\textbf{Moritz}}
    & \multicolumn{3}{c}{\textbf{Zhou2016}}
    & \multirow{2.5}{*}{\textbf{Mean}} \\
  \cmidrule(lr){2-4} \cmidrule(lr){5-7} \cmidrule(lr){8-10}
  &  \rotatebox{0}{\textbf{M5}} 
  &  \rotatebox{0}{\textbf{M6}} 
  &  \rotatebox{0}{\textbf{M7}}
  &  \rotatebox{0}{\textbf{M5}} 
  &  \rotatebox{0}{\textbf{M6}} 
  &  \rotatebox{0}{\textbf{M7}}
  &  \rotatebox{0}{\textbf{M5}} 
  &  \rotatebox{0}{\textbf{M6}} 
  &  \rotatebox{0}{\textbf{M7}}
    &  \\
  \midrule
  Baseline
    & 83.56 & 79.01 & 79.91
    & 68.17 & 62.40 & 57.33
    & 74.35 & 74.55 & 73.50 & 72.53 \\
 \midrule
    AdaBN 
    & 83.56\addGre{~~~0.00} & 79.97\addred{↑0.96} & 79.91\addGre{~~~0.00}
    & 68.17\addGre{~~~0.00} & 63.38\addred{↑0.98} & 58.23\addred{↑0.90}
    & 70.45\add{↓3.90} & 71.39\add{↓3.16} & 76.25\addred{↑2.75} & 72.37\add{↓0.16} \\
  AdaBN (W/ EA)
    & 83.41\add{↓0.15} & 78.02\add{↓0.99} & 80.21\addred{↑0.30}
    & \cellcolor{LightRed}\textbf{69.37}\addred{↑1.20} & 63.37\addred{↑0.97} & 58.27\addred{↑0.94}
    & 73.45\add{↓0.90} & 73.39\add{↓1.16} & 73.80\addred{↑0.30} & 72.59\addred{↑0.06} \\
  Tent 
    & 82.79\add{↓0.77} & \cellcolor{LightRed}\textbf{81.79}\addred{↑2.78} & 80.16\addred{↑0.25}
    & 67.50\add{↓0.67} & 65.60\addred{↑3.20} & 52.73\add{↓4.60}
    & 71.38\add{↓2.97} & 69.92\add{↓4.63} & 64.83\add{↓8.67} & 70.74\add{↓1.79} \\
  Tent (W/ EA)
    & 81.89\add{↓1.67} & 80.89\addred{↑1.88} & 80.18\addred{↑0.27}
    & 66.36\add{↓1.81} & 63.43\addred{↑1.03} & 58.73\addred{↑1.40}
    & 68.27\add{↓6.08} & 66.83\add{↓7.72} & 60.61\add{↓12.89} & 69.69\add{↓2.84} \\
  T3A 
    & 49.97\add{↓33.59} & 72.43\add{↓6.58} & 79.28\add{↓0.63}
    & 39.90\add{↓28.27} & 60.40\add{↓2.00} & 57.67\addred{↑0.34}
    & 48.16\add{↓26.19} & 69.39\add{↓5.16} & 76.04\addred{↑2.54} & 61.47\add{↓11.06} \\
  T3A (W/ EA)
    & 50.01\add{↓33.55} & 76.49\add{↓2.52} & 79.78\add{↓0.13}
    & 40.17\add{↓28.00} & 60.17\add{↓2.23} & 57.63\addred{↑0.30}
    & 48.71\add{↓25.64} & 69.38\add{↓5.17} & 73.14\add{↓0.36} & 61.72\add{↓10.81} \\
  T-TIME$^{\star}$
    & \cellcolor{shallowBlue}84.66\addred{↑1.10} & 78.68\add{↓0.33} &  81.38\addred{↑1.47}
    & 67.73\add{↓0.44} & \cellcolor{LightRed}\textbf{67.49}\addred{↑5.09} & \cellcolor{shallowBlue}61.96\addred{↑4.63}
    & \cellcolor{shallowBlue}81.02\addred{↑6.67} &  76.85\addred{↑2.30} &  76.50\addred{↑3.00} & 75.14\addred{↑2.61} \\
  T-TIME$^{\star\star}$
    & 84.36\addred{↑0.80} & 79.85\addred{↑0.84} & \cellcolor{shallowBlue}81.45\addred{↑1.54}
    & 68.00\add{↓0.17} & \cellcolor{shallowBlue}67.39\addred{↑4.99} & 59.56\addred{↑2.23}
    & 80.76\addred{↑6.41} & \cellcolor{shallowBlue}80.50\addred{↑5.95} & \cellcolor{shallowBlue}77.79\addred{↑4.29} & \cellcolor{shallowBlue}75.52\addred{↑2.99} \\
  \midrule
  Ours
    & \cellcolor{LightRed}\textbf{84.95}\addred{↑1.39} & \cellcolor{shallowBlue}80.96\addred{↑1.95} & \cellcolor{LightRed}\textbf{81.41}\addred{↑1.50}
    & \cellcolor{shallowBlue}68.93\addred{↑0.76} & 67.10\addred{↑4.70} & \cellcolor{LightRed}\textbf{62.27}\addred{↑4.94}
    & \cellcolor{LightRed}\textbf{81.68}\addred{↑7.33} & \cellcolor{LightRed}\textbf{81.04}\addred{↑6.49} & \cellcolor{LightRed}\textbf{77.21}\addred{↑3.71} & \cellcolor{LightRed}\textbf{76.17}\addred{↑3.64} \\
  \bottomrule
  \end{tabular}
    \begin{tablenotes}
      \footnotesize
          \item[1] M5–M7 denote EEGNet, DeepConvNet, and EEGConformer, respectively. 
          \item[2] The results compare the proposed algorithm with four TTA methods: AdaBN, Tent, T3A, and T-TIME ($^{\star}$single-trial update, $^{\star\star}$batch update with batch-size = 8). 
          \item[3] Listed are the performance of AdaBN, Tent, and T3A with Euclidean alignment (W/ EA).
          \item[4] The best result per column is in\bgred{red}; the second best is in\bgblue{blue}.  
  \end{tablenotes}
  \end{threeparttable}
\end{table*}

\subsection{Backbone Networks}
 
\subsubsection{SSVEP Backbone Networks}
\begin{itemize}
    \item 
    \textit{CNN-M, CNN-C}~(\textbf{M1, M2})\footnote{\url{https://github.com/aaravindravi/Brain-computer-interfaces}}~\cite{ravi2020comparing}: compact CNNs that operate on the features derived by fast Fourier transform and learn representations via convolution and pooling.
    \item
    \textit{FBCNN-M, FBCNN-C}~(\textbf{M3, M4})\footnote{\url{https://github.com/tianwangchn/}}~\cite{zhao2021filter}: FBCNN variants that decompose EEG into multiple sub-bands and use parallel CNN branches to learn frequency-specific representations.
\end{itemize}

\subsubsection{Motor Imagery Backbone Networks}
\begin{itemize}
    \item \textit{EEGNet}~(\textbf{M5})\footnote{\url{https://github.com/vlawhern/arl-eegmodels/blob/master/EEGModels.py}}~\cite{lawhern2018eegnet}: EEGNet is a compact CNN that applies depth-wise and point-wise convolutions to decouple spatial and temporal filtering.
    \item
    \textit{DeepConvNet}~(\textbf{M6})\footnote{\url{https://github.com/robintibor/braindecode}}~\cite{schirrmeister2017deep}: DeepConvNet is a deep CNN that stacks temporal convolution with spatial convolution across channels to learn temporal–spatial representations.
    \item
    \textit{EEGConformer}~(\textbf{M7})\footnote{\url{https://github.com/eeyhsong/EEG-Conformer}}~\cite{song2022eeg}: EEGConformer is designed based on a hybrid convolution transformer architecture to capture long range temporal dependencies.
\end{itemize}

\subsection{Test-Time Adaptation Algorithms for Comparison}
\begin{itemize}
    \item 
    \textit{AdaBN}~(ICLR 2017)~\cite{li2016revisiting}: AdaBN adapts by recomputing batch normalization statistics on target-domain data and substituting the source-domain statistics to align feature distributions. 
    \item
    \textit{Tent}~(ICLR 2021)~\cite{wang2020tent}\footnote{\url{https://github.com/DequanWang/tent}}~: Tent minimizes prediction Shannon entropy and updates the affine parameters of the batch normalization layers during inference, enabling label-free adaptation at test time. 
    \item
    \textit{T3A}~(NeurIPS 2021)~\cite{iwasawa2021test}\footnote{\url{https://github.com/matsuolab/T3A}}~: T3A replaces classifier weights with class prototypes computed from confident test-time predictions.
    \item
    \textit{T-TIME}~(TBME 2023)~\cite{li2023t}\footnote{\url{https://github.com/sylyoung/DeepTransferEEG}}: T-TIME performs calibration-free EEG adaptation via conditional-entropy minimization and adaptive marginal-distribution regularization over sliding test batches, using an ensemble for pseudo-labeling.
\end{itemize}

\subsection{Datasets}

Experiments cover both SSVEP and motor imagery paradigms with five datasets. Two publicly available SSVEP datasets are used (35 and 10 subjects), and three motor imagery datasets are used (105, 10, and 4 subjects).

\subsubsection{SSVEP datasets}

\begin{itemize}
  \item \textit{Benchmark~\cite{wang2016benchmark}}: 35 subjects are recorded with a 64-channel EEG system at 1000\,Hz (downsampled to 250\,Hz). Each subject completes 240 trials (6 blocks \(\times\) 40), using visual stimuli from 40 flicker-frequency classes (8–15.8\,Hz in 0.2\,Hz steps). Each trial lasts 5\,s with signals from all 64 electrodes.
  \item \textit{12JFPM~\cite{nakanishi2015comparison}}: 10 subjects are recorded using eight occipital electrodes at 2048\,Hz (downsampled to 256\,Hz). Each subject completes 180 trials (15 blocks \(\times\) 12), with 12 joint frequency–phase classes: frequencies 9.25–14.75\,Hz in 0.5\,Hz steps and phases started from 0 with an interval of 0.5$\pi$. Each trial lasts 5\,s.    
\end{itemize}
 
\subsubsection{Motor imagery datasets}
\begin{itemize}              
\item \textit{PhysioNet~\cite{schalk2004bci2000}}: 109 subjects are recorded by a 64-channel EEG system at 160\,Hz. Tasks include imagining opening/closing the left/right fists. Each trial comprises 2\,s preparation, 4\,s imagery, and 2\,s rest. After excluding four subjects for data quality issues, 105 subjects remain.
\item \textit{Moritz~\cite{grosse2009beamforming}}: 10 subjects are recorded with a 128-channel EEG system at 500\,Hz, totaling 300 trials. Tasks include imagining finger moving and gripping an object. Each trial includes 3\,s preparation and 7\,s imagery; rest duration is not reported.
\item \textit{Zhou2016~\cite{zhou2016fully}}: 4 subjects are recorded with a 14-channel EEG system at 250\,Hz, totaling 250 trials. Tasks include left- and right-hand motor imagery. Each trial includes 1\,s preparation, 5\,s imagery, and 4\,s rest.
\end{itemize}

\begin{table}[!t]\small
\renewcommand{\arraystretch}{1.2}
\centering
\caption{average accuracy gains (\%) of different TTA methods across backbone on different datasets.}
\label{tab:improvements_datasets}
\begin{threeparttable}
  \begin{tabular}{%
     p{1.0cm}<{\centering}  
     p{0.85cm}<{\centering}   
     p{0.7cm}<{\centering}   
     p{0.9cm}<{\centering}     
     p{1.0cm}<{\centering}   
     p{1.1cm}<{\centering}   
     p{0.4cm}<{\centering}   
  }
  \toprule
  \multirow{2.4}{*}{Dataset}
    & \multicolumn{6}{c}{\textbf{TTA Method}} \\ 
  \cmidrule(lr){2-7}
    & \shift{-0.2em}{\textbf{AdaBN}}
    & \shift{-0.1em}{\textbf{Tent}}
    & \shift{-0.3em}{\textbf{T3A}}
    & \shift{-0.4em}{\textbf{T-TIME$^\star$}}
    & \shift{-0.2em}{\textbf{T-TIME$^{\star\star}$}}
    & \shift{0.1em}{\textbf{Ours}} \\
  \midrule
  Benchmark &3.73&2.56&-5.83 &6.26 &  \cellcolor{shallowBlue}6.50 &
     \cellcolor{LightRed}\textbf{7.58} \\
  12JFPM\,   &  1.05 & -1.11 &   1.27 &\cellcolor{shallowBlue}1.58 &  0.62 &
     \cellcolor{LightRed}\textbf{2.23} \\
  PhysioNet\,  & -0.28 &  0.16 & -12.07 &  0.75 &  \cellcolor{shallowBlue}1.06 &
     \cellcolor{LightRed}\textbf{1.61} \\
  Moritz\,    &  1.04 & 0.21 &  -9.98 &  \cellcolor{shallowBlue}3.09 &  2.35 &
    \cellcolor{LightRed} \textbf{3.47} \\
  Zhou2016\,       & -0.59 & -8.90 & -10.39 &  3.99 &  \cellcolor{shallowBlue}5.55 &
    \cellcolor{LightRed} \textbf{5.84} \\
  \bottomrule
  \end{tabular}
      \begin{tablenotes}
      \footnotesize
          \item[1] The best result per row is in\bgred{red}; the second best is in\bgblue{blue}.
          \item[2] $^{\star}$single-trial update, $^{\star\star}$batch update with batch-size = 8. 
  \end{tablenotes}
\end{threeparttable}
\end{table}

\begin{table}[!t]\small
\renewcommand{\arraystretch}{1.2}
\centering
\caption{average accuracy gains (\%) of different TTA methods across datasets for different backbones.}
\label{tab:improvements_backbones}
\begin{threeparttable}
  \begin{tabular}{%
     p{1.0cm}<{\centering}  
     p{0.7cm}<{\centering}   
     p{0.7cm}<{\centering}   
     p{0.9cm}<{\centering}     
     p{1.0cm}<{\centering}   
     p{1.1cm}<{\centering}   
     p{0.4cm}<{\centering}   
  }
  \toprule
  \multirow{2.4}{*}{Backbone}
    & \multicolumn{6}{c}{\textbf{TTA Method}} \\
  \cmidrule(lr){2-7}

 & \shift{0.2em}{\raisebox{-0\height}{\rotatebox{0}{\textbf{AdaBN}}}}
& \shift{0.2em}{\raisebox{-0\height}{\rotatebox{0}{\textbf{Tent}}}}
& \shift{-0.1em}{\raisebox{-0\height}{\rotatebox{0}{\textbf{T3A}}}}
& \shift{-0.3em}{\raisebox{-0\height}{\rotatebox{0}{\textbf{T-TIME}$^{\star}$}}}
& \shift{0.0em}{\raisebox{-0\height}{\rotatebox{0}{\textbf{T-TIME}$^{\star\star}$}}}
& \shift{0.0em}{\raisebox{-0\height}{\rotatebox{0}{\textbf{Ours}}}} \\

  \midrule
 M1        & 3.96 & 2.51 &  -0.10 & 4.87 & \cellcolor{shallowBlue}5.36 &  \cellcolor{LightRed}\textbf{6.14} \\
M2        & 0.03 & -2.67 & -9.91 & \cellcolor{shallowBlue}1.57 & 1.02 &  \cellcolor{LightRed}\textbf{3.38} \\
M3     & 4.73 & 4.00 & 2.82 & 5.90 & \cellcolor{shallowBlue}6.23 &  \cellcolor{LightRed}\textbf{6.88} \\
 M4   & 0.84 & -0.94 & -1.93 & \cellcolor{LightRed}\textbf{3.35} & 1.64 &  \cellcolor{shallowBlue}3.20 \\
  M5       & 0.05 & -3.19 & -29.06 & \cellcolor{shallowBlue}2.44 & 2.35 & \cellcolor{LightRed} \textbf{3.16} \\
  M6  & -0.39 & -1.60 & -3.31 & 2.35 & \cellcolor{shallowBlue}3.93 &  \cellcolor{LightRed}\textbf{4.38} \\
  M7 & 0.51 & -3.74 & -0.06 & \cellcolor{shallowBlue}3.03 & 2.69 & \cellcolor{LightRed} \textbf{3.38} \\
  \bottomrule
  \end{tabular}
      \begin{tablenotes}
      \footnotesize
          \item[1] M1–M7 denote CNN-M, CNN-C, FBCNN-M, FBCNN-C, EEGNet, DeepConvNet, and EEGConformer respectively. 
          \item[2] The best result per row is in\bgred{red}; the second best is in\bgblue{blue}. 
  \end{tablenotes}
\end{threeparttable}
\end{table}

\begin{figure*}
    \centering
    \includegraphics[width=1\linewidth]{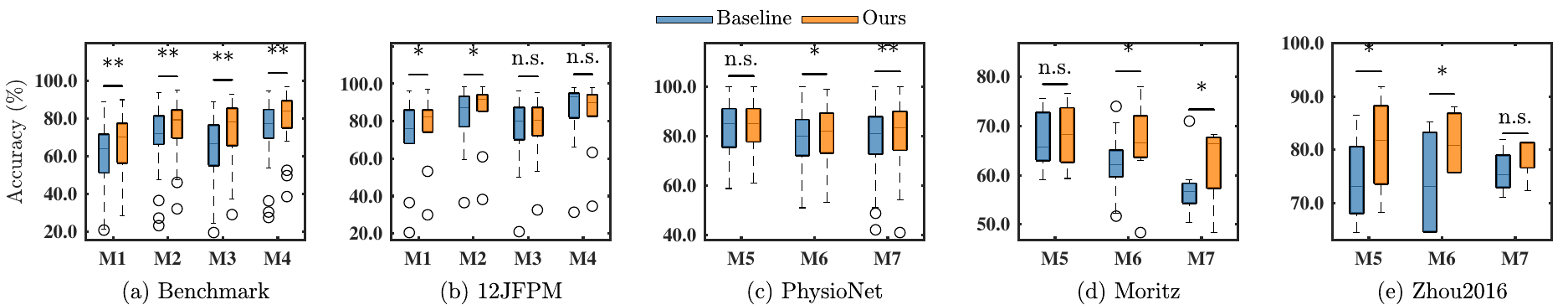}
     \caption{The average accuracy (\%) of different methods across five datasets, where blue bars represent Baseline and orange bars represent the proposed algorithm (Ours). Statistical significance between methods is assessed using paired t-tests, with $^{*}$ indicating significance levels: $^{*}$ for $\mathrm{p} < 0.05$, and $^{**}$ for $\mathrm{p} < 0.01$, and n.s. indicates non-significant. Subfigures (a)–(e) correspond to the Benchmark~\cite{wang2016benchmark}, 12JFPM~\cite{nakanishi2015comparison}, PhysioNet~\cite{schalk2004bci2000}, Moritz~\cite{grosse2009beamforming}, and Zhou2016~\cite{zhou2016fully} datasets.}

    \label{fig:box}
\end{figure*}

\subsection{Experimental Results}

\subsubsection{Generalization across Multiple Datasets}

The proposed algorithm demonstrates consistent gains across datasets, 7.58\%, 2.23\%, 1.61\%, 3.47\%, and 5.84\% on the Benchmark~\cite{wang2016benchmark}, 12JFPM~\cite{nakanishi2015comparison}, PhysioNet~\cite{schalk2004bci2000}, Moritz~\cite{grosse2009beamforming}, and Zhou2016~\cite{zhou2016fully} datasets, respectively. Based on paired t-tests in Fig.~\ref{fig:box}, twelve out of seventeen baselines method comparisons are statistically significant (p $<$ 0.05), while the non-significant cases are concentrated in 12JFPM (FBCNN-M and FBCNN-C) and a few motor imagery settings (PhysioNet–EEGNet, Moritz–EEGNet, Zhou2016–EEGConformer), likely reflecting smaller data sizes. A more detailed analysis is carried out based on the scale of datasets, on the relatively large-scale datasets Benchmark, PhysioNet, Moritz, the average improvement over the corresponding baselines is 4.22\%; on smaller-scale datasets 12JFPM, Zhou2016, the average improvement is 4.04\%.

These results suggest that comparable gains across both scales indicate that the method provides complementary benefits at test time. Although larger cohorts and more trials yield stronger base models, there still remains substantial headroom. These findings support applicability across datasets of varying scales, and underscore the utility of test-time adaptation for EEG decoding.
 
\begin{figure}[!t]
    \centering
    \includegraphics[width=1\linewidth]{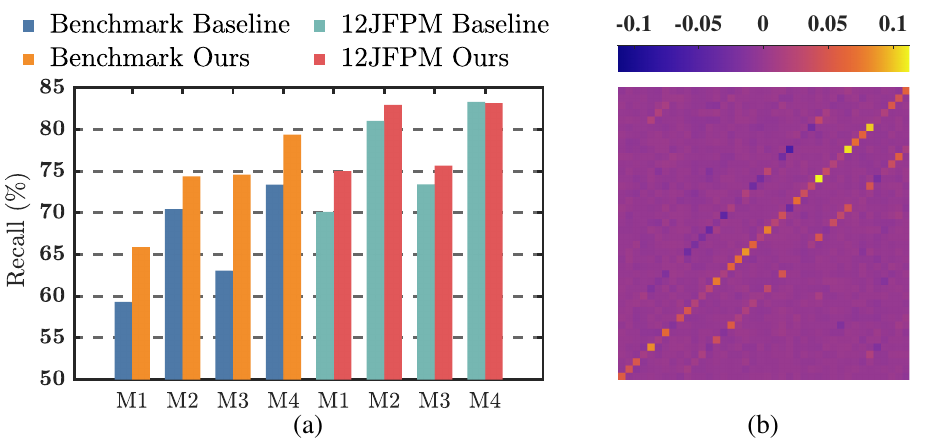}
    \caption{(a) Comparison of recall rates of different categories between the SSVEP backbone model baseline and the proposed algorithm on the Benchmark and 12JFPM datasets (\%); (b) is the difference confusion matrix on the 40 category Benchmark dataset using the CNN-C model, obtained by subtracting the baseline CNN-C results from those of the proposed method.}
    \label{fig:recall-label}
\end{figure}

\subsubsection{Different Paradigms Generalizability}

The SSVEP paradigm relies on steady-state oscillations in the occipital cortex induced by periodic visual stimulation, which appear as peaks at the stimulus frequencies and their harmonics in spectral analyses~\cite{vialatte2010steady}. By contrast, the motor imagery paradigm elicits ERD/ERS in the \(\mu\) (8–13\,Hz) and \(\beta\) (13–30\,Hz) bands over sensorimotor areas, manifested as task-dependent decreases or increases in band power~\cite{neuper2006erd}.

Although these paradigms differ fundamentally in neural mechanisms and signal characteristics leading to different feature distributions for online adaptation, the proposed algorithm exhibits cross-paradigm generalization. Compared to the corresponding baseline, average accuracy increases by 4.90\% on SSVEP datasets~\cite{wang2016benchmark,nakanishi2015comparison} and by 3.64\% on motor imagery datasets~\cite{schalk2004bci2000,grosse2009beamforming,zhou2016fully}, indicating robust performance across both paradigms. Consistently, paired t-tests on Fig.~\ref{fig:box} indicate that most model paradigm combinations achieve statistically significant gains (p $<$ 0.05), with the few non-significant outcomes mainly arising in smaller datasets or where baseline performance is already high.

These findings suggest that the proposed TTA algorithm can accommodate heterogeneous EEG characteristics and is applicable across BCI paradigms.

\subsubsection{Plug-and-Play Capability}


The proposed algorithm is plug-and-play across diverse EEG backbones. When applied during offline training and test-time adaptation, it improves inter-subject decoding performance without modifying the underlying classifier architectures.

As summarized in Table~\ref{tab:improvements_backbones}, consistent gains are observed across backbones. For SSVEP, the mean accuracy of CNN-M and CNN-C~\cite{ravi2020comparing} increases by approximately 4.75\%, and that of FBCNN-M and FBCNN-C~\cite{zhao2021filter} by approximately 5.04\%. For motor imagery, the average improvements are 3.16\% for EEGNet~\cite{lawhern2018eegnet}, 4.38\% for DeepConvNet~\cite{schirrmeister2017deep}, and 3.38\% for EEGConformer~\cite{song2022eeg}. Despite architectural differences, all backbones benefit, indicating that the method maintains plug-and-play compatibility.



\subsubsection{Performance Gains Across All Categories Within a Dataset}

The proposed algorithm yields consistent class-wise improvements. Fig.~\ref{fig:recall-label} compares recall between the baselines and the proposed method on the Benchmark dataset (40 classes)~\cite{wang2016benchmark} and the 12JFPM dataset (12 classes)~\cite{nakanishi2015comparison}. On the Benchmark dataset, average recall increases by approximately 7\% across the four decoder backbones; comparable gains are observed on the 12JFPM dataset. 

Additionally, Fig.~\ref{fig:recall-label} (b) presents the difference confusion matrix on the 40 category Benchmark dataset~\cite{wang2016benchmark} using the CNN-C model~\cite{ravi2020comparing}, obtained by subtracting the baseline CNN-C results from those of the proposed method. The brighter diagonal band indicates higher correct-class rates across most categories, rather than improvements confined to a few classes. These class-wise improvements reduce the need for class-specific calibration at deployment, thereby lowering system complexity and enhancing reliability.

\subsubsection{Comparison with Test-Time Adaptation Methods}


The proposed algorithm shows the best performance compared with existing TTA methods across multiple datasets and baselines in most cases (12/17), as summarized in Table~\ref{tab:combined} and Table~\ref{tab:mi_combined}. Three observations arise. First, each stage of the dual-stage alignment is critical. Methods lacking data-level alignment, such as AdaBN and Tent exhibit lower decoding performance, with the proposed approach exceeding them by 5.11–6.72\% on SSVEP and 3.80–5.43\% on motor imagery. Second, in the absence of representation-level alignment (e.g., T-TIME), performance saturates: the proposed method surpasses T-TIME$^{\star\star}$ by 0.98\% on SSVEP and 1.03\% on motor imagery, indicating that omitting representation-level alignment becomes a bottleneck for the decoder. Third, T3A without both data- and representation-level alignment and relying solely on classifier prototype adjustment can even fall below the source-only baseline on both SSVEP and motor imagery.

For methods that initially lack data-level alignment, introducing data-level TTA yields further gains. The data-aligned variants AdaBN~(W/EA), Tent~(W/EA), and T3A~(W/EA) achieve higher accuracies than their original counterparts, with additional improvements of +2.60\%, +2.55\%, and +2.64\% on SSVEP and +0.22\%, $-1.05$\%, and +0.25\% on motor imagery, respectively (see Table~\ref{tab:combined} and Table~\ref{tab:mi_combined}). These results underscore the importance of alignment at the data level. Even with data-level alignment, these methods remain behind the proposed algorithm. Relative to the baseline, the proposed method delivers gains of +4.90\% on SSVEP and +3.64\% on motor imagery and surpasses the second best method by +0.98\% and +0.65\%, respectively (Table~\ref{tab:combined}; Table~\ref{tab:mi_combined}). Beyond dual-stage alignment, the strategy for updating the decoder is therefore pivotal; the proposed self-supervised loss is effective for online decoder adaptation.

\begin{table}[!t]\small
\renewcommand{\arraystretch}{1.2} 
\centering
\caption{Accuracy of ablation study on the Benchmark dataset (\%).}  
\label{tab:ablation}
\begin{threeparttable}
\begin{tabular}{
  >{\centering\arraybackslash}p{0.5cm}
  >{\centering\arraybackslash}p{0.5cm}
  >{\centering\arraybackslash}p{0.5cm}
  >{\centering\arraybackslash}p{1.0cm}
  >{\centering\arraybackslash}p{1.0cm}
  >{\centering\arraybackslash}p{1.0cm}
  >{\centering\arraybackslash}p{1.0cm}
}
\toprule
\multirow{1}{*}{\textbf{EA}}
  & \multirow{1}{*}{\textbf{BN}}
  & \multirow{1}{*}{\textbf{$\mathcal{L}$}}
  & \textbf{M1} & \textbf{M2} & \textbf{M3} & \textbf{M4} \\
\midrule
 
    $\times$ & $\checkmark$ & $\checkmark$
      & 60.33 & 70.33 & 62.58 & 73.08 \\
    $\checkmark$ & $\times$ & $\checkmark$
      & \cellcolor{shallowBlue}66.07 
      & \cellcolor{shallowBlue}72.48 
      & 73.03 & 78.34 \\
    $\checkmark$ & $\checkmark$ & $\times$
      & 64.84 & 70.71 
      & 73.43 
      & \cellcolor{shallowBlue}78.59 \\
    $\times$ & $\times$ & $\checkmark$
      & 61.90 & 71.50 
      & 53.35 & 68.16 \\
    $\times$ & $\checkmark$ & $\times$
      & 57.72 & 68.37 
      & 60.93 & 71.82 \\
    $\checkmark$ & $\times$ & $\times$
      & 65.05 & 70.69 
      & \cellcolor{shallowBlue}74.01 
      & 76.88 \\
    \midrule
 
    $\times$ & $\times$ & $\times$
      & 59.68 & 70.15 
      & 63.27 & 73.52 \\
 
    $\checkmark$ & $\checkmark$ & $\checkmark$
      & \cellcolor{LightRed}\textbf{67.18} 
      & \cellcolor{LightRed}\textbf{74.46} 
      & \cellcolor{LightRed}\textbf{74.98} 
      & \cellcolor{LightRed}\textbf{80.31} \\
\bottomrule
\end{tabular}
    \begin{tablenotes}
      \footnotesize
          \item[1] M1–M4 denote CNN-M, CNN-C, FBCNN-M, and FBCNN-C. 
          \item[2] The best result per column is in\bgred{red}; the second best is in\bgblue{blue}.   
          \item[3] EA: Euclidean Alignment; BN: Batch Normalization Statistics Update; $\mathcal{L}$: Self-Supervised Loss.
          \item[4] $\checkmark$ and $\times$ indicate whether the corresponding component is enabled or disabled, respectively.
  \end{tablenotes}
\end{threeparttable}
\end{table}

\begin{table}[!t]\small
\renewcommand{\arraystretch}{1.2}
\centering
\caption{Batch-size sensitivity under different component settings on the Benchmark dataset (Accuracy,~\%).}
\label{tab:batch_ablation_merged}
\begin{threeparttable}
\begin{tabular}{
  >{\centering\arraybackslash}p{0.2cm}
  >{\centering\arraybackslash}p{0.2cm}
  >{\centering\arraybackslash}p{0.2cm}
  >{\centering\arraybackslash}p{0.2cm}
  >{\centering\arraybackslash}p{1.1cm}
  >{\centering\arraybackslash}p{1.1cm}
  >{\centering\arraybackslash}p{1.1cm}
  >{\centering\arraybackslash}p{1.1cm}
}
\toprule
\textbf{EA} & \textbf{BN} & \textbf{$\mathcal{L}$} & \textbf{BS} & \textbf{M1} & \textbf{M2} & \textbf{M3} & \textbf{M4} \\
\midrule
$\checkmark$ & $\checkmark$ & $\checkmark$ & 1 & 67.18 & 74.46 & 74.98 & 80.31 \\
$\checkmark$ & $\checkmark$ & $\checkmark$ & 2 & \cellcolor{shallowBlue}67.32 & 74.55 & 74.90 & 80.13 \\
$\checkmark$ & $\checkmark$ & $\checkmark$ & 4 & \cellcolor{LightRed}\textbf{67.35} & 74.45 & 74.95 & 80.19 \\
$\checkmark$ & $\checkmark$ & $\checkmark$ & 6 & 67.27 & \cellcolor{shallowBlue}74.56 & \cellcolor{LightRed}\textbf{75.11} & \cellcolor{shallowBlue}80.90 \\
$\checkmark$ & $\checkmark$ & $\checkmark$ & 8 & 67.21 & \cellcolor{LightRed}\textbf{74.59} & \cellcolor{shallowBlue}75.07 & \cellcolor{LightRed}\textbf{81.30} \\
\midrule
$\checkmark$ & $\times$ & $\checkmark$ & 1 & \cellcolor{shallowBlue}66.07 & 72.48 & 73.03 & 78.34 \\
$\checkmark$ & $\times$ & $\checkmark$ & 2 & 65.89 & 72.51 & \cellcolor{LightRed}\textbf{73.33} & 78.52 \\
$\checkmark$ & $\times$ & $\checkmark$ & 4 & 66.07 & \cellcolor{shallowBlue}72.69 & 73.26 & \cellcolor{LightRed}\textbf{78.84} \\
$\checkmark$ & $\times$ & $\checkmark$ & 6 & \cellcolor{LightRed}\textbf{66.09} & \cellcolor{LightRed}\textbf{72.70} & 73.29 & \cellcolor{shallowBlue}78.83 \\
$\checkmark$ & $\times$ & $\checkmark$ & 8 & 66.06 & 72.68 & \cellcolor{shallowBlue}73.30 & 78.82 \\
\midrule
$\checkmark$ & $\checkmark$ & $\times$ & 1 & 64.84 & 70.71 & 73.43 & 78.59 \\
$\checkmark$ & $\checkmark$ & $\times$ & 2 & 64.98 & 70.74 & 73.55 & 78.61 \\
$\checkmark$ & $\checkmark$ & $\times$ & 4 & 65.14 & 70.75 & 73.62 & \cellcolor{LightRed}\textbf{78.65} \\
$\checkmark$ & $\checkmark$ & $\times$ & 6 & \cellcolor{LightRed}\textbf{65.30} & \cellcolor{shallowBlue}70.77 & \cellcolor{shallowBlue}73.63 & \cellcolor{shallowBlue}78.64 \\
$\checkmark$ & $\checkmark$ & $\times$ & 8 & \cellcolor{shallowBlue}65.17 & \cellcolor{LightRed}\textbf{70.81} & \cellcolor{LightRed}\textbf{73.65} & 78.63 \\
\bottomrule
\end{tabular}
\begin{tablenotes}
  \footnotesize
  \item[1] BS: Batch-Size. EA: Euclidean Alignment; BN: Batch normalization statistics update; $\mathcal{L}$: Self-supervised loss. 
  \item[2] All results are on the benchmark SSVEP dataset. The best result per column is in\bgred{red}; the second best is in\bgblue{blue}.
  \item[3] $\checkmark$ and $\times$ indicate whether the corresponding component is enabled or disabled, respectively. 
\end{tablenotes}
\end{threeparttable}
\end{table}

\subsection{Ablation Study}

\subsubsection{Dual‑Stage Alignment}
As shown in Table~\ref{tab:ablation}, an ablation study on the Benchmark dataset using CNN-M, CNN-C~\cite{ravi2020comparing}, FBCNN-M, and FBCNN-C~\cite{zhao2021filter} quantifies the contribution of each alignment stage. Removing the full dual-stage alignment (data-level and representation-level alignment) reduces the mean classification accuracy by 10.5\%. Considering each stage separately, ablating Euclidean alignment in data-level leads to a 7.6\% decrease, while ablating the batch normalization statistics update in the representation stage yields a 1.7\% decrease. These results indicate that both stages contribute individually, with Euclidean alignment exerting the larger effect. Moreover, the sum of the individual drops (7.6\%+1.7\%=9.3\%) is slightly smaller than the joint ablation (10.5\%), suggesting the dual-stage alignment is largely complementary, with separable contributions in the two stages.

\subsubsection{Self-Supervised Loss}
The self-supervised objective is employed to adapt the method to target-domain data during test time. As shown in Table~\ref{tab:ablation}, removing this objective reduces mean accuracy by 2.2\%, indicating the contribution to online adaptation through updates to decoder parameters. Moreover, an interesting interaction effect is observed: the self-supervised objective is effective only in the presence of the dual-stage alignment, particularly the data-level Euclidean alignment. Without alignment, applying the objective can further degrade performance. This outcome is expected because the self-supervised objective leverages model predictions derived from the training distribution to supervise unlabeled test data; when source–target shift is large, such supervision can reinforce mis-calibrated predictions. By reducing this shift, alignment brings test data closer to the training domain, yielding more reliable pseudo-targets and thereby improving performance.

\subsubsection{A Better Trade-Off Between Accuracy and Online Responsiveness}

The proposed algorithm supports single-trial updates and attains a favorable trade-off between decoding performance and online responsiveness. Online responsiveness is governed by the calibration time, defined as the latency before the first update while trials are accumulated. Given typical trial durations approximately 1-2~s/trial for SSVEP and 4-5~s/trial for motor imagery aggregating multiple trials substantially increases this latency. With all components enabled, increasing the batch-size does not improve accuracy. As reported in Table~\ref{tab:batch_ablation_merged}, across all SSVEP decoders, raising the batch-size from 1 to 8 yields at most an average accuracy gain of only 0.22\%. Accordingly, the algorithm achieves single-trial updates that maintain accuracy while reducing calibration latency. 

To examine the influence of key components (Table~\ref{tab:batch_ablation_merged}), removing either the batch normalization statistics update or the self-supervised loss produces accuracy lower than the single-trial, all components enabled configuration, even when multiple trials are accumulated. This indicates that performance gains are driven by the integrity and coordination of the core components rather than by a larger batch-size.

Component wise sensitivity to batch-size is further assessed without ablation by increasing the batch-size for only one component, either the batch normalization statistics update or the self-supervised loss (Table~\ref{tab:batchsize_loss_bn_joint}). The benefit is minimal: average accuracy increases by merely +0.07\% and +0.01\%, respectively. These results suggest that the algorithm is well adapted to single-trial updates and exhibits low sensitivity to batch-size, an effect plausibly attributable to the dual-stage alignment, which improves the estimation accuracy and robustness of distribution statistics and thereby reduces dependence on larger batches.

\begin{table}[!t]\small
\renewcommand{\arraystretch}{1.2}
\centering
\caption{Batch-size sensitivity of self-supervised loss updates and batch normalization statistics updates with both components enabled on the Benchmark dataset. (Accuracy,~\%)}
\label{tab:batchsize_loss_bn_joint}
\begin{threeparttable}
  \begin{tabular}{%
     p{0.8cm}<{\centering}  
     p{1.0cm}<{\centering}  
     p{0.5cm}<{\centering}  
     p{0.8cm}<{\centering}  
     p{0.8cm}<{\centering}  
     p{0.8cm}<{\centering}  
     p{0.8cm}<{\centering}  
  }
  \toprule
  \textbf{$\mathcal{L}$-BS} & \textbf{BN-BS} & \textbf{BS} & \textbf{M1} & \textbf{M2} & \textbf{M3} & \textbf{M4} \\
  \midrule
  $\checkmark$ & $\times$ & 1 & 67.18 & 74.46 & \cellcolor{LightRed}\textbf{74.98} & 80.31 \\
  $\checkmark$ & $\times$ & 2 & \cellcolor{LightRed}\textbf{67.44} & \cellcolor{shallowBlue}74.49 & 74.78 & 80.32 \\
  $\checkmark$ & $\times$ & 4 & 67.37 & \cellcolor{LightRed}\textbf{74.57} & 74.79 & 80.46 \\
  $\checkmark$ & $\times$ & 6 & 67.38 & \cellcolor{LightRed}\textbf{74.57} & \cellcolor{shallowBlue}\textbf{74.87} & \cellcolor{LightRed}\textbf{80.52} \\
  $\checkmark$ & $\times$ & 8 & \cellcolor{shallowBlue}67.42 & \cellcolor{LightRed}\textbf{74.57} & 74.84 & \cellcolor{shallowBlue}80.48 \\
  \midrule
  $\times$ & $\checkmark$ & 1 & 67.18 & \cellcolor{LightRed}\textbf{74.46} & 74.98 & \cellcolor{LightRed}\textbf{80.31} \\
  $\times$ & $\checkmark$ & 2 & \cellcolor{LightRed}\textbf{67.41} & 74.35 & 75.00 & \cellcolor{shallowBlue}80.29 \\
  $\times$ & $\checkmark$ & 4 & 67.32 & 74.43 & 74.80 & 80.20 \\
  $\times$ & $\checkmark$ & 6 & \cellcolor{shallowBlue}67.40 & 74.42 & \cellcolor{shallowBlue}75.01 & 80.27 \\
  $\times$ & $\checkmark$ & 8 & 67.36 & \cellcolor{shallowBlue}74.45 & \cellcolor{LightRed}\textbf{75.05} & 80.11 \\
  \bottomrule
  \end{tabular}
  \begin{tablenotes}
    \footnotesize
    \item[1] The best result per column is in\bgred{red}; the second best is in\bgblue{blue}. 
    \item[2] Both components are enabled. A $\checkmark$ in $\mathcal{L}$-BS or BN-BS indicates that the corresponding component is updated with the batch-size of that row, while the other component is fixed to single-trial (BS=1). 
    \item[3] $\checkmark$ and $\times$ indicate whether the corresponding component is enabled or disabled, respectively.
  \end{tablenotes}
\end{threeparttable}
\end{table}

Overall, with all components enabled, single-trial updates substantially shorten the calibration time while providing decoding performance comparable to that of multi-trial updates. The favorable performance therefore results from the joint effect of all components together with the single-trial update scheme, yielding a better trade-off between decoding performance and online efficiency.

\begin{figure*}[!t]
    \centering
    \includegraphics[width=1\linewidth]{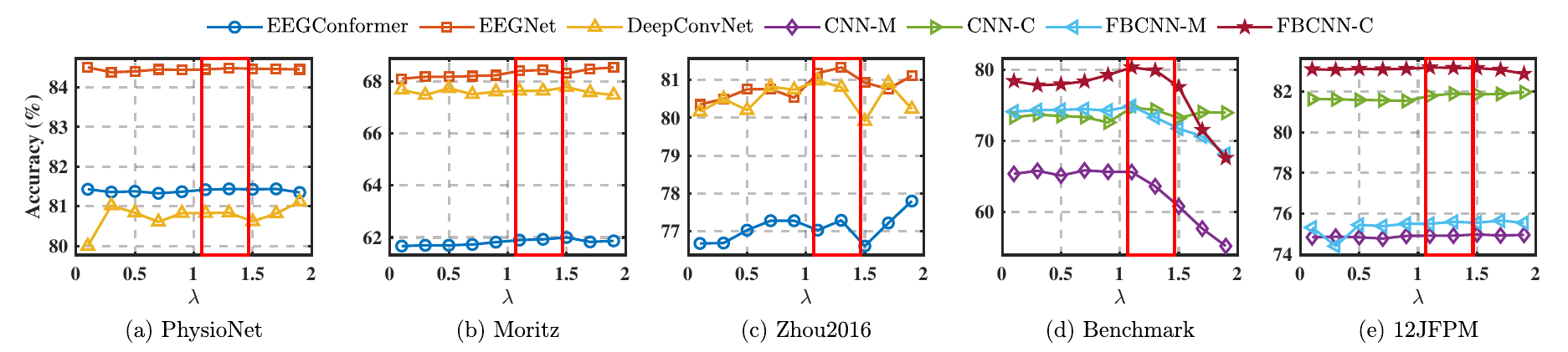}
     \caption{The effect of $\lambda$ on the average accuracy of various decoder backbones across five datasets. (a)–(e) correspond to the PhysioNet~\cite{schalk2004bci2000}, Moritz~\cite{grosse2009beamforming}, Zhou2016~\cite{zhou2016fully}, Benchmark~\cite{wang2016benchmark}, and 12JFPM~\cite{nakanishi2015comparison} datasets. The red box indicates the selected range of $\lambda$, which is approximately 1.1–1.4.}
    \label{fig:lam_use}
\end{figure*}
\section{Discussion}

\subsection{Calibration of the Estimation Bias of the Cross-Entropy Loss with $\lambda$}

Because \(\mathcal{L}_{\mathrm{CE}}\) is approximated by \(\hat{\mathcal{L}}^{\mathrm{val}}_{\mathrm{CE}}\), the proxy can exhibit systematic bias and therefore requires calibration via the coefficient \(\lambda\). Analysis shown in Appendix~A establishes:
\begin{itemize}
  \item If \(A_{\mathrm{val}} > A_{\mathrm{test}}\), then \(\hat{\mathcal{L}}^{\mathrm{val}}_{\mathrm{CE}}\) underestimates \(\mathcal{L}_{\mathrm{CE}}\).
  \item If \(A_{\mathrm{val}} < A_{\mathrm{test}}\), then \(\hat{\mathcal{L}}^{\mathrm{val}}_{\mathrm{CE}}\) overestimates \(\mathcal{L}_{\mathrm{CE}}\).
\end{itemize}

Accordingly, a grid search over \(\lambda \in [0.1, 1.9]\) was conducted across backbones and datasets (Fig.~\ref{fig:lam_use}), yielding peak performance for \(\lambda \approx 1.1\text{–}1.4\). For example, on the Benchmark dataset with CNN-M, \(A_{\mathrm{val}} \approx 66\%\) and \(A_{\mathrm{test}} \approx 65\%\), consistent with the case \(A_{\mathrm{val}} > A_{\mathrm{test}}\) where the proxy underestimates \(\mathcal{L}_{\mathrm{CE}}\); the selected \(\lambda\) range compensates for this bias, in line with the analysis in Appendix~A.

\subsection{Input Data- and Representation-level Alignment}
 
Online Euclidean alignment and batch normalization statistics updating play complementary roles in mitigating inter-subject distribution shifts. The former provides initial calibration of the covariance structure, whereas the latter continuously adjusts the feature distribution as new data arrive, thereby ensuring stability of the representation space. The combination of the two not only enhances feature separability but also allows the classifier to maintain high generalization performance on unseen subjects. Ablation results further indicate that removing either alignment mechanism leads to noticeable degradation, highlighting the necessity of dual-stage alignment for effective online adaptation.

\begin{figure}
    \centering
    \includegraphics[width=1\linewidth]{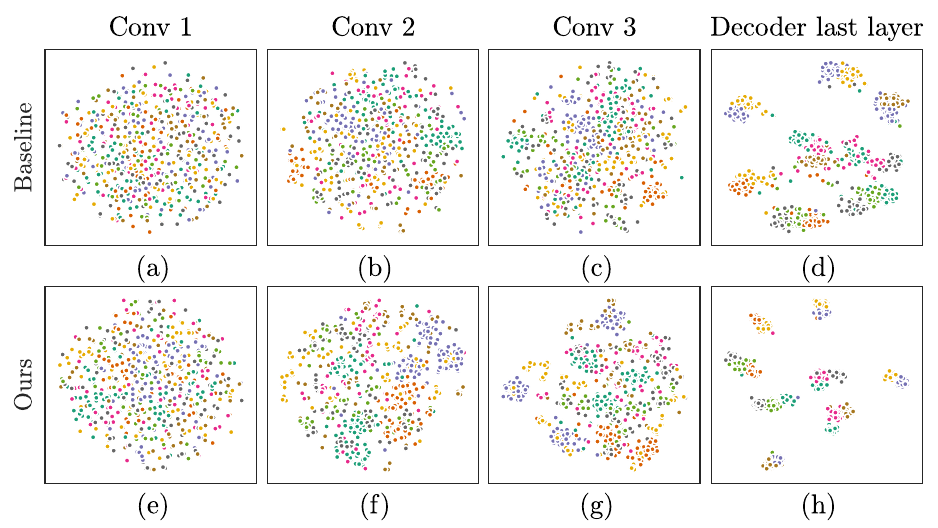}
    \caption{t-SNE visualizations of FBCNN-C~\cite{zhao2021filter} on subject S2 from the Benchmark dataset~\cite{wang2016benchmark}, illustrating the feature distributions of the first 20 categories across different network layers. (a)–(d) show the baseline t-SNE results of Conv1 to Conv3 and the decoder last layer without applying Euclidean alignment and updating batch normalization statistics, while (e)–(h) show the corresponding visualizations of the same layers after applying the proposed algorithm.}
    \label{fig:combined}
\end{figure}

In addition, Fig.~\ref{fig:combined} presents the t-SNE visualizations of FBCNN-C~\cite{zhao2021filter} on subject S2 from the Benchmark dataset~\cite{wang2016benchmark}, showing the feature distributions of the first 20 categories across different network layers, from Conv1 to Conv3 as well as the decoder last layer. Without any alignment, feature clusters from different trials are relatively dispersed, and mixing occurs between different classes (Fig.~\ref{fig:combined} (a) to (d)). After applying online Euclidean alignment and updating batch normalization statistics, the clusters at all layers become noticeably more compact, and class separability is improved.

Quantitative results in Table~\ref{tab:distances} present intra- to inter-class distance ratios obtained from the t-SNE embeddings of different layers. The values consistently decrease when comparing the baselines to the proposed method, with ratios reduced from 4.80 to 3.12 at Conv1, from 1.42 to 1.11 at Conv2, from 1.62 to 0.55 at Conv3, and from 0.22 to 0.08 at last layer of the decoder. A smaller ratio indicates that trials belonging to the same class are more compact while different classes are more separated, thereby reflecting stronger discriminative representations. These findings confirm that the proposed method enhances feature separability across layers, providing a more favorable representation space for reliable classification.

\begin{table}[!t] \small
\renewcommand{\arraystretch}{1.2}
\centering

\caption{The ratio of intra-class to inter-class distance within each convolutional layer and the last layer of the decoder on FBCNN-C~\cite{zhao2021filter}. Smaller values indicate that trials of the same class are more compact while different classes are farther apart, i.e., better class separability.}
\label{tab:distances}
\begin{threeparttable}
\begin{tabular}{
  >{\centering\arraybackslash}p{1.0cm}
  >{\centering\arraybackslash}p{3.0cm}
  >{\centering\arraybackslash}p{3.0cm}
  >{\centering\arraybackslash}p{3.0cm}
}
\toprule
\multirow{2.4}{*}{\textbf{Layer}} 
  & \multicolumn{2}{c}{\textbf{Intra/Inter Ratio}} \\
\cmidrule(lr){2-3}
  & \textbf{Baseline} & \textbf{Ours} \\
\midrule
Conv1 & 4.80 & \textbf{3.12} \\
Conv2 & 1.42 & \textbf{1.11} \\
Conv3 & 1.62 & \textbf{0.55} \\
Last   & 0.22 & \textbf{0.08} \\
\bottomrule
\end{tabular}
\begin{tablenotes}
  \footnotesize
  \item[1] Ratios are computed as the mean intra-class distance divided by the mean inter-class distance, based on the 2D t-SNE embeddings of each layer. 
  \item[2] A smaller ratio indicates more compact intra-class clusters and larger inter-class separation, thus reflecting better class separability.
\end{tablenotes}
\end{threeparttable}
\end{table}

\subsection{Performance Improvement across Trials}

\begin{figure}
    \centering
    \includegraphics[width=1\linewidth]{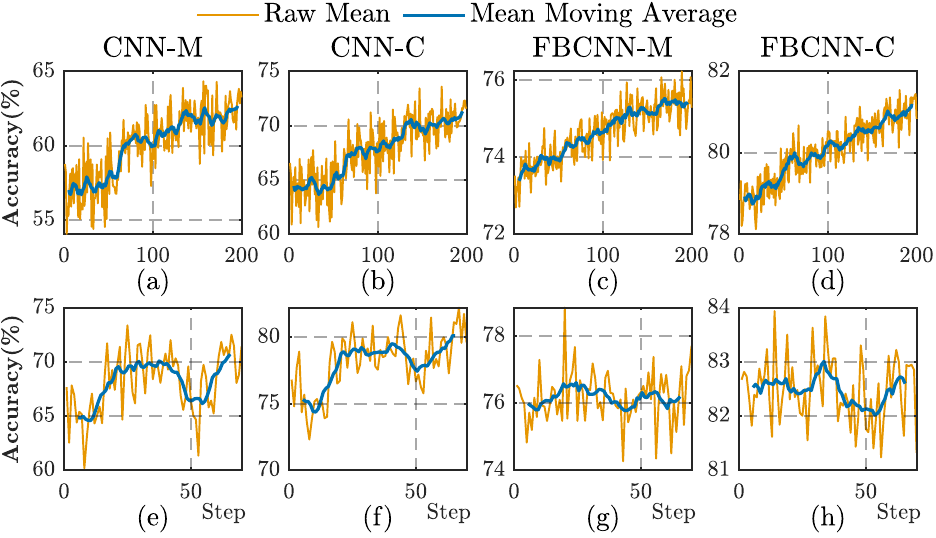}
  \caption{Experimental results on the Benchmark~\cite{wang2016benchmark} and 12JFPM~\cite{nakanishi2015comparison} datasets under the LOSO cross-validation protocol. For each subject, the first half of the trials is used as the TTA test set, and the second half is reserved as a fixed validation set that does not participate in any model update. The Raw Mean denotes the accuracy curve averaged over all subjects, while Mean Moving Average refers to the smoothed curve obtained by applying a moving average with a window size of 10. Subfigures (a)–(d) show the Benchmark dataset and (e)–(h) show 12JFPM, both with CNN-M, CNN-C, FBCNN-M, and FBCNN-C.}
    \label{fig:Subjects_window_smoothed_combined_fourplot}
\end{figure}



\begin{figure}[!t]
    \centering
    \includegraphics[width=1\linewidth]{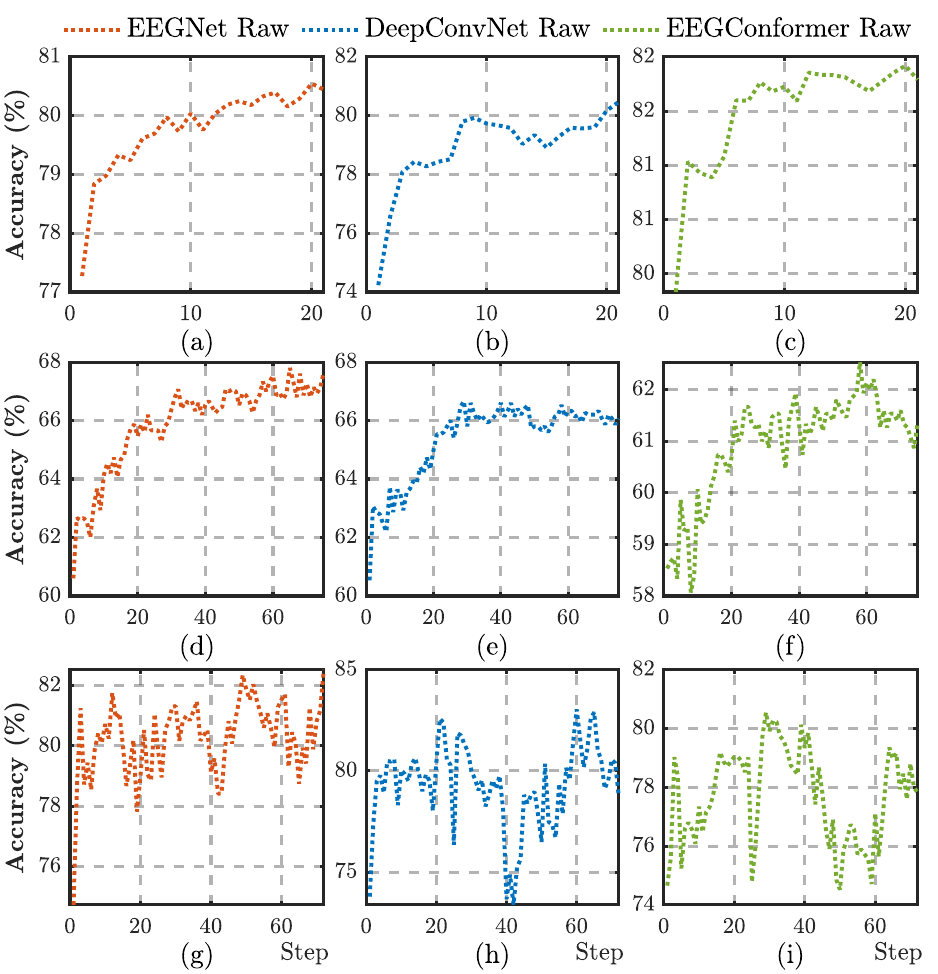}
\caption{Experimental results on the motor imagery datasets under the LOSO cross-validation protocol. For each subject, the first half of the trials is used as the test set, while the second half serves as a fixed validation set that does not participate in any model update. Specifically, (a)–(c) show the TTA curves on the PhysioNet~\cite{schalk2004bci2000}, (d)–(f) correspond to the Moritz~\cite{grosse2009beamforming}, and (g)–(i) present the results on the Zhou2016~\cite{zhou2016fully}.}
    \label{fig:placeholder}
\end{figure}
  
As shown in Fig.~\ref{fig:Subjects_window_smoothed_combined_fourplot} and Fig.~\ref{fig:placeholder}, consistent patterns are observed across datasets. On relatively larger-scale datasets (e.g., Benchmark~\cite{wang2016benchmark}, PhysioNet~\cite{schalk2004bci2000}, and Moritz~\cite{grosse2009beamforming}), the average accuracy increases rapidly at the early stage of TTA and then gradually stabilizes. In the SSVEP paradigm on the Benchmark dataset, the accuracies of all decoders increase approximately linearly with the number of TTA steps; in contrast, in the motor imagery paradigm on PhysioNet and Moritz, the accuracies of different decoders follow a logarithmic growth and gradually saturate. These results indicate that larger datasets provide richer target distributions, which support stable performance gains during the online stage. By contrast, relatively smaller-scale datasets (e.g., 12JFPM~\cite{nakanishi2015comparison} and Zhou2016~\cite{zhou2016fully}) display more pronounced fluctuations; although neither a linear increasing trend nor a logarithmic trend is evident, the accuracies still improve over the baseline, indicating that a certain degree of adaptation remains feasible.

These observations indicate that dataset scale is one of the key factors affecting TTA effectiveness. More extensive utilization of large-scale datasets can potentially lead to more stable improvements in practical applications. This highlights a direction for future research, where training and adaptation on larger inter-subject or different paradigm datasets may further enhance the generalization capability and practical value of EEG decoding models in diverse environments.

\section{Conclusion}

This study presents a plug-and-play online adaptation algorithm for EEG-based BCI systems. The algorithm incorporates a dual-stage alignment procedure and a self-supervised loss. The dual-stage alignment calibrates target-domain data in both the EEG data space and the representation space. The self-supervised loss employs soft pseudo-label based cross-entropy adjusted by Shannon entropy. Extensive experiments demonstrate that the algorithm integrates seamlessly with diverse EEG decoder backbones and paradigms. The encouraging results pave the way for online adaptation in EEG decoding, expanding the applicability of non-invasive BCI systems. This capability benefits practical BCI deployment by reducing calibration requirements and enhancing robustness of EEG decoders. Future studies will concentrate on deploying the proposed algorithm in BCI systems to verify its feasibility in real-world applications.

\appendices 
\section{Using the accuracy of the validation dataset  as the probability value for the predicted class}
Under the assumption introduced in \eqref{eq:pseudo_label}, using $A_{\mathrm{val}}$ as the predicted class’s probability involves the following considerations, the Expected Calibration Error (ECE) \cite{guo2017calibration} is defined as
\begin{equation}
\label{eq:ece_def}
\mathrm{ECE} = \sum_{m=1}^{M}\frac{|B_{m}|}{N_{\mathrm{val}}}\bigl|\mathrm{acc}(B_{m})-\mathrm{conf}(B_{m})\bigr|
\end{equation}
In this formula, \(M = 10\) specifies the number of confidence intervals, \(B_{m}\) denotes the set of validation samples whose predicted confidence lies in interval \(m\), \(\lvert B_{m}\rvert\) is the size of \(B_{m}\), \(N_{\mathrm{val}}\) is the total number of validation samples, \(\mathrm{acc}(B_{m})\) is the actual classification accuracy of samples in \(B_{m}\), and \(\mathrm{conf}(B_{m})\) is the mean predicted confidence of those samples. For example, the CNN-M~\cite{ravi2020comparing} model from the Benchmark dataset~\cite{wang2016benchmark} achieved an average ECE of approximately 0.02 on the validation set. The low ECE shows that predicted confidences closely match actual accuracies. Based on the model's demonstrated ECE value, the \(\hat{p}^\mathrm{val}(\hat{Y}^\mathrm{test}_i = k\mid x^\mathrm{test}_i)\) is used as an approximation for \(\hat{p}^\mathrm{test}(\hat{Y}^\mathrm{test}_i = k \mid x^\mathrm{test}_i)\) .

\section{Bias Analysis under Soft Label Assumption}

This appendix analyzes the estimation bias of the soft pseudo-label cross‐entropy loss when both pseudo labels and true labels are modeled as Bernoulli soft distributions.
\subsection{Soft Label Distributions}
Define two soft label distributions over classes:

\begin{itemize}
\item {\bf Soft Pseudo-label distribution} \(\hat{p}^\mathrm{test}\bigl(\hat{Y}^\mathrm{test}_i=k\mid x^\mathrm{test}_i\bigr)\) based on test accuracy \(A_{\mathrm{test}}\)
\begin{equation}
  \label{eq:appendix_true}
  \hat{p}^\mathrm{test}\bigl(\hat{Y}^\mathrm{test}_i=k\mid x^\mathrm{test}_i\bigr)
  =\begin{cases}
    A_{\mathrm{test}}, & k = y^\mathrm{test}_i, \\[4pt]
    \dfrac{1 - A_{\mathrm{test}}}{K-1}, & k \neq y^\mathrm{test}_i
  \end{cases}
\end{equation}
where $\hat{p}^\mathrm{test}\bigl(\hat{Y}^\mathrm{test}_i=k\mid x^\mathrm{test}_i\bigr)$ based on final test accuracy \(A_{\mathrm{test}}\) (unknown until after testing), and \(\hat{p}^\mathrm{val}\bigl(\hat{Y}^\mathrm{test}_i=k\mid x^\mathrm{test}_i\bigr)\) based on validation set accuracy \(A_{\mathrm{val}}\):
\begin{equation}
  \label{eq:appendix_pseudo}
  \hat{p}^\mathrm{val}\bigl(\hat{Y}^\mathrm{test}_i=k\mid x^\mathrm{test}_i\bigr)
  =\begin{cases}
    A_{\mathrm{val}}, & k = y_i^*, \\[4pt]
    \dfrac{1 - A_{\mathrm{val}}}{K-1}, & k \neq y_i^*
  \end{cases}
  \end{equation}
\end{itemize}

\subsection{Closed-form Cross-Entropy Loss}
Define the positive-class loss
\[
\ell_i^+ = -\log q\bigl(\hat {Y}^\mathrm{test}_i = y_i^* \mid x^\mathrm{test}_i)
\]
and the average negative-class loss
\[
\ell_i^- = -\frac{1}{K-1}
            \sum_{k \neq y_i^*}
            \log q\bigl(\hat{Y}^\mathrm{test}_i = k \mid x^\mathrm{test}_i)
\]
so that \(\ell_i^+ < \ell_i^-\).  Then the soft pseudo-label cross-entropy is
\begin{equation}
\label{eq:appendix_Lce}
\begin{aligned}
\hat{\mathcal{L}}_{{\mathrm{CE}}}^\mathrm{val}
&=-\sum_{k=1}^K
  \hat{p}^\mathrm{val}(\hat{Y}^\mathrm{test}_i=k\mid x_i)\,
  \log q\bigl(\hat Y^\mathrm{test}_i=k\mid x_i)\\
&= A_{\mathrm{val}}\,\ell_i^+ \;+\;(1-A_{\mathrm{val}})\,\ell_i^-
\end{aligned}
\end{equation}

Similarly, using the true-label distribution ${p}$ gives
\begin{equation}
\label{eq:appendix_Lce_true}
\mathcal{L}_{{\mathrm{CE}}}^\mathrm{test}
= A_{\mathrm{test}}\,\ell_i^+ \;+\;(1-A_{\mathrm{test}})\,\ell_i^-
\end{equation}

\subsection{Monotonicity and Bias}
Differentiating \eqref{eq:appendix_Lce} with respect to \(A_{\mathrm{val}}\) yields
\[
\frac{\partial \hat{\mathcal{L}}^\mathrm{val}_{{\mathrm{CE}}}}{\partial A_{\mathrm{val}}}
= \ell_i^+ - \ell_i^- < 0
\]
so \(\hat{\mathcal{L}}_{{\mathrm{CE}}}^\mathrm{val}\) decreases as \(A_{\mathrm{val}}\) increases.  Subtracting \eqref{eq:appendix_Lce_true} from \eqref{eq:appendix_Lce} gives
\begin{equation}
\label{eq:appendix_diff}
\hat{\mathcal{L}}_{{\mathrm{CE}}}^\mathrm{val}
- {\mathcal{L}}_{{\mathrm{CE}}}^\mathrm{test}
= (A_{\mathrm{val}} - A_{\mathrm{test}})\,(\ell_i^+ - \ell_i^-)
\end{equation}
Since \(\ell_i^+ - \ell_i^-<0\):
\begin{itemize}
  \item if \(A_{\mathrm{val}} > A_{\mathrm{test}}\), then 
        \(\hat{\mathcal{L}}_{{\mathrm{CE}}}^\mathrm{val} < \mathcal{L}_{{\mathrm{CE}}}^\mathrm{test}\), 
        indicating underestimation of the true loss;
  \item if \(A_{\mathrm{val}} < A_{\mathrm{test}}\), then 
        \(\hat{\mathcal{L}}_{{\mathrm{CE}}}^\mathrm{val} > \mathcal{L}_{{\mathrm{CE}}}^\mathrm{test}\), 
        indicating overestimation of the true loss.
\end{itemize}

\subsection{Implication for \(\lambda\)}
The loss is
\[
\mathcal{L}_{\mathrm{test}}
= \mathcal{L}_{\mathrm{Ent}}
+ \lambda\bigl(\hat{\mathcal{L}}_{{\mathrm{CE}}}^\mathrm{val} - \mathcal{L}_{\mathrm{Ent}}\bigr)
\]
To correct the bias in \(\hat{\mathcal{L}}^\mathrm{val}_{{\mathrm{CE}}}\):
\[
\begin{cases}
\lambda > 1, & \text{if }A_{\mathrm{val}} > A_{\mathrm{test}},\\
\lambda < 1, & \text{if }A_{\mathrm{val}} < A_{\mathrm{test}}
\end{cases}
\]

\subsection{Experimental Consistency}
In the experiments, the pseudo-label accuracy is \(A_{\mathrm{val}} = 0.66\) and the final test accuracy is \(A_{\mathrm{test}} = 0.65\). Since \(A_{\mathrm{val}} > A_{\mathrm{test}}\), theory predicts \(\lambda>1\). 
Grid search on validation data found \(\lambda \approx 1.1\) to yield the best performance, consistent with this analysis.

{\small
\bibliographystyle{IEEEtran}
\bibliography{mybib.bib}
}

\end{document}